# On the electrical conductivity of plasmas and metals

B J B Crowley[1,2,3]

[1]Department of Physics, University of Oxford, Parks Road, Oxford OX1 3PU, UK
[2]AWE PLC, Reading RG7 4PR, UK

[3]Email: *basil.crowley@physics.ox.ac.uk*

Date: 28 September 2015

Methods for modelling the electrical conductivity of dense plasmas and liquid metals, based upon the well-known Ziman formula, are reviewed from a general perspective, and some earlier inconsistencies relating to its application to finite temperature systems are resolved. A general formula for the conductivity of a Lorentzian two-component plasma in thermal equilibrium is derived from the Lenard-Balescu collision integral in which both energy and momentum exchange between ions and electrons are accounted for. This formula is used as a basis for some generalizations of the Ziman formula, which apply to plasmas of arbitrary degeneracy over a much wider range of conditions. These formulae implicitly include the collective motions of the ions, but neglect the collective motions of the electrons. Detailed consideration of the latter shows that they generally have a small effect on the conductivity. Conditions for the validity of the Ziman formula are derived. The extension of the general theory to arbitrarily low temperatures, where the ion dynamics become dominated by collective effects, in which dynamical ion correlations need to be taken into account, is shown to lead to the well-known Bloch formula.
Consideration is given to non-Lorentzian plasmas by explicitly accounting for electron-electron collisions. Corrections to the Lorentzian model in the form of a power series in $1/Z$ are derived.

**PACS**: **51.50.+v; 52.20.-j;** 52.20.Fs; 52.25.Fi; **72.15.-v;** 72.15.Cz;  72.15.Lh

**Keywords:** Electrical conductivity of plasmas and metals, Lorentzian plasmas, Ziman conductivity, Gruneisen-Bloch conductivity, electron-ion collision frequency, relaxation time, collision integral, Coulomb logarithm, non-Lorentzian plasmas.

This page is intentionally left blank

# CONTENTS

# 1 INTRODUCTION

## 1.1 Introduction

The electrical conductivity of disordered Coulomb systems (plasmas and liquid metals) is a property of considerable theoretical and experimental importance. Measurements of the electrical conductivity of bulk matter can be carried out directly and yield information about the charge carrier density and the static correlations between the atomic ions, and hence about the structure of the material. These correlations are represented in the well-known Ziman formula for the conductivity of a liquid metal by the ion static structure factor, $S_{ii}(\mathbf{q})$ which is the Fourier transform of the instantaneous two-particle correlation function.

From a more practical point of view, the modelling of the electrical conductivity of plasmas formed by electrical discharges, such as Z-pinches and capillary discharges, is important for understanding the dynamics of the discharge process.

Collisional processes, including charged particle transport and scattering, can be modelled in terms of dynamic structure factor(s), $S(\mathbf{q},\omega)$, which, in the linear regime, are related, via the *fluctuation dissipation theorem*, to the longitudinal dielectric function(s) $\varepsilon(\mathbf{q},\omega)$. This offers a more consistent and powerful approach to treating these problems, in which a full description of the system dynamics is contained in the dielectric function.

## 1.2 Summary

This paper comprises a pedagogical review of the Ziman formula, and various generalizations of it, for the static conductivity of liquid metals and plasmas, with the aim of deriving a single formula whose validity extends across the widest possible range of regimes.

Generalized formulae, along with conditions for their validity, are derived using a common underlying model based upon a finite-temperature Boltzmann equation treatment of the electrons. This is initially combined with a static approximation to the electron-ion collision integral (Landau), in which the ions remain stationary. For the main study, a fully dynamic approximation based upon the linearized Lenard-Balescu collision integral, which properly accounts for energy transfer to the ions, is used.

In the process, the Faber-Ziman formula is rederived to include a treatment of the prefactor, thus properly extending the formula to regimes of arbitrary degeneracy, ranging from fully degenerate (where the original Ziman formula applies) to the Maxwell-Boltzmann limit. In the regimes of low-degeneracy and weak coupling, the Maxwell-Boltzmann result, in the appropriate classical or semi-classical limits, reverts to the well-known formulae for gases and low density plasmas given by Spitzer and Landau.

The overall treatment is simpler than many of those given elsewhere and removes inconsistencies previously noted, and incorrectly ascribed to different levels of approximation.

The standard Ziman formula is shown not to extend to temperatures below the Debye temperature(s) corresponding to the collective modes of the ion subsystem. A correction to the formula that, in principle, allows generalization to arbitrarily low temperatures is proposed. When this correction is applied, the description reduces to the Gruneisen-Bloch formula for the resistivity of a normal metal, but with extensions to higher densities.

In this way, a unification of different models for the electrical conductivity of plasmas and metals covering a wide range of different regimes is achieved.

Consideration is given to the generalization to non-Lorentzian plasmas, one in which the additional effects of electron-electron collisions are accounted for. Corrections, in the form of an expansion in powers of $1/Z$ around the Lorentzian limit, are derived.

## 2 CONDUCTIVITY OF A LORENTZIAN PLASMA

### 2.1 Basic formulae

The electrical conductivity of a Lorentzian plasma (one in which the effect of electron-electron collisions is disregarded) comprising electrons and ions is given, in principle exactly, by

$$\sigma = \frac{2e^2}{3m_e T_e} \frac{1}{\mathfrak{V}} \sum_\beta \varepsilon_\beta \tau_\beta p_\beta q_\beta$$

$$\equiv \frac{2e^2 n_e}{3m_e T_B} \langle\langle \varepsilon\tau \rangle\rangle$$

(1)

(see equation (158) in APPENDIX A) where the summation is over continuum electron quasiparticle states $\beta$ occupying volume $\mathfrak{V}$ and having energies, $\varepsilon_\beta = \varepsilon(\mathbf{k}_\beta)$, ($\varepsilon(\mathbf{k}) = k^2/2m_e$), and relaxation times, $\tau_\beta$, The notation $\langle\langle \ \rangle\rangle$ is explained in APPENDIX C. Here $p_\beta$ denotes the Fermi-Dirac distribution

$$p_\beta = \frac{1}{1+\exp(\varepsilon_\beta/T_e - \eta)}$$

(2)

and $q_\beta = 1 - p_\beta$. The effective temperature, $T_B$, is defined in the first instance, by [1] $T_B = T_e / \langle q \rangle$, where $\langle \; \rangle$ denotes an average taken over the distribution (2). Transforming the summation in equation (1) to an integral over wavenumber, using

$$\frac{1}{\mathfrak{V}} \sum_\beta = \frac{g}{(2\pi)^3} \int d^3 \mathbf{k} \tag{3}$$

(where $g = 2$ for electrons), yields, making use of (2),

$$\sigma = \frac{e^2}{6\pi^3 m_e T_e} \int \varepsilon \tau p q \, d^3 \mathbf{k}$$

$$= -\frac{e^2}{3\pi^2 m_e} \int_0^\infty \frac{\partial p}{\partial k} \tau k^3 \, dk \tag{4}$$

## 3  THE STATIC APPROXIMATION - THE ZIMAN CONDUCTIVITY FORMULA

For weak quasi-elastic collisions in a static potential, the relaxation time, which is the reciprocal of the collision frequency, $\nu$, is given as follows

$$\frac{1}{\tau(k)} = \nu(k) = \frac{n_i m_e}{4\pi k^3} \int_0^{2k} |\tilde{V}(q)|^2 S_{ii}(q) q^3 \, dq \tag{5}$$

(See APPENDIX A, equation (170).) where $\tilde{V}(q) = V(q) / \varepsilon_e(q, 0)$ is the statically-screened electron-ion potential and $S_{ii}(q)$ is the ion static structure factor [2], [3], [4], [5], [6]. Note that the static screening is realised by the electronic part $\varepsilon_e$ of the dielectric function only. This is a significant point, one which we shall return to later. Equations (4) and (5) agree with equations 9.146 of ref.[7] (apart from a missing sign in one of the latter – see also equation 2.224 of ref [7]).

In the degenerate limit, in which $\langle\langle \varepsilon \tau \rangle\rangle \to T_F \tau_F$ where $T_F = k_F^2 / 2m_e$ is the *Fermi temperature* and $\tau_F = \tau(k_F)$, one obtains the standard Ziman formula [8]:

$$\sigma = \frac{2e^2 n_e}{3 m_e T_B} T_F \tau_F = \frac{e^2 n_e}{m_e} \left( \frac{4\pi k_F^3}{n_i m_e} \right) \left[ \int_0^{2k_F} |\tilde{V}(q)|^2 S_{ii}(q) q^3 \, dq \right]^{-1}$$

$$= \frac{12\pi^3 n_e Z e^2}{m_e^2} \left[ \int_0^{2k_F} |\tilde{V}(q)|^2 S_{ii}(q) q^3 \, dq \right]^{-1} \tag{6}$$

where $Z = n_e / n_i$. In the following, we discuss some further generalizations of this formula.

### 3.1 Generalizations of the Ziman formula

We now approach the resistivity from the standard formula

$$\sigma^{-1} = \frac{m_e \bar{v}}{n_e e^2} \tag{7}$$

making use of the (approximate) formula, APPENDIX A, equation (177),

$$\bar{v} = \frac{3T_B}{2} \frac{\langle\langle \varepsilon v \rangle\rangle}{\langle\langle \varepsilon^{5/2} \rangle\rangle \langle\langle \varepsilon^{-1/2} \rangle\rangle} \tag{8}$$

for the mean collision frequency, where (cf. equation (1)),

$$\langle\langle \varepsilon v \rangle\rangle = \frac{T_B}{T_e} \frac{1}{n_e \mathfrak{V}} \sum_\beta \varepsilon_\beta v_\beta p_\beta q_\beta \tag{9}$$

And where, referring to (178)-(179) in APPENDIX A,

$$\langle\langle \varepsilon^{5/2} \rangle\rangle \langle\langle \varepsilon^{-1/2} \rangle\rangle = \frac{3}{2} \frac{T_B^2}{\pi^2 n_e} \frac{\langle k^3 \rangle}{1+e^{-\eta}} \tag{10}$$

This yields

$$\bar{v} = \frac{\pi^2 n_e (1+e^{-\eta})}{T_B \langle k^3 \rangle} \langle\langle \varepsilon v \rangle\rangle$$

$$= \frac{\pi^2 (1+e^{-\eta})}{T_e \langle k^3 \rangle} \frac{1}{\mathfrak{V}} \sum_\beta \varepsilon_\beta v_\beta p_\beta q_\beta \tag{11}$$

Transforming the summation in equation (11) to an integral over wavenumber by means of (3), with $g=2$, in the manner of (4), and subsequently integrating by parts, assuming $\lim_{k \to 0}(k^3 v(k)) = 0$, yields

$$\frac{1}{\mathfrak{V}} \sum_\beta \varepsilon_\beta v_\beta p_\beta q_\beta = \frac{2}{(2\pi)^3} \int \varepsilon v p q \, d^3 \mathbf{k}$$

$$= -\frac{T_e}{2\pi^2} \int_0^\infty \frac{\partial p}{\partial k} v k^3 \, dk \tag{12}$$

$$= \frac{T_e}{2\pi^2} \int_0^\infty p(k) \frac{\partial}{\partial k}(v k^3) \, dk$$

where, from (5),

$$\frac{\partial}{\partial k}\left(\nu k^{3}\right) = \frac{n_{i}m_{e}}{2\pi}\left|\tilde{V}(2k)\right|^{2} S_{ii}(2k)(2k)^{3} \tag{13}$$

Substituting (13) into (12) yields, without further approximation,

$$\frac{1}{\mathfrak{V}}\sum_{\beta}\varepsilon_{\beta}\nu_{\beta}p_{\beta}q_{\beta} = \frac{n_{i}m_{e}T_{e}}{(2\pi)^{3}}\int_{0}^{\infty} p\left(\tfrac{1}{2}q\right)\left|\tilde{V}(q)\right|^{2} S_{ii}(q)q^{3}\,\mathrm{d}q \tag{14}$$

whereupon the mean collision frequency given by (11) becomes

$$\overline{\nu} = \frac{n_{i}m_{e}}{8\pi}\frac{\left(1+\mathrm{e}^{-\eta}\right)}{\left\langle k^{3}\right\rangle}\int_{0}^{\infty} p\left(\tfrac{1}{2}q\right)\left|\tilde{V}(q)\right|^{2} S_{ii}(q)q^{3}\,\mathrm{d}q \tag{15}$$

and the resistivity (7) is

$$\sigma^{-1} = \frac{m_{e}^{2}}{8\pi Ze^{2}}\frac{\left(1+\mathrm{e}^{-\eta}\right)}{\left\langle k^{3}\right\rangle}\int_{0}^{\infty} p\left(\tfrac{1}{2}q\right)\left|\tilde{V}(q)\right|^{2} S_{ii}(q)q^{3}\,\mathrm{d}q \tag{16}$$

Apart from a factor of $3\pi^{2}n_{e}\left(1+\mathrm{e}^{-\eta}\right)/2\left\langle k^{3}\right\rangle$, which is unity in the degenerate limit, and equal to $3\pi/32$ in the non-degenerate (Boltzmann) limit [9], this is the same as the generalization [10], of the Ziman formula, as given at equation (2.227) in [7]. However, the factor is important. For a start, it removes the inconsistency identified in [7], between the formula given there and equation (1). Since equation (16) is here derived directly from (1), there is no inconsistency in the result given above.

In the degenerate limit, $\left\langle k^{3}\right\rangle \to \tfrac{1}{2}k_{F}^{3} = \tfrac{3}{2}\pi^{2}n_{e}$ and $p(k)=1:\ k<k_{F};\ =0:\ k>k_{F}$, and (16) reduces exactly to the standard low-temperature form of the Ziman formula (6).

Equation (16) can also be evaluated in the non-degenerate (Maxwell-Boltzmann) limit ($\eta \ll 0$) for which

$$p(k) \to \exp\left(\eta - k^{2}/2m_{e}T_{e}\right) \tag{17}$$

and

$$\left\langle k^{3}\right\rangle \to \frac{4}{\sqrt{\pi}}\left(2m_{e}T_{e}\right)^{3/2} \tag{18}$$

to yield

$$\sigma^{-1} = \frac{1}{32\sqrt{\pi}Ze^{2}}\frac{m_{e}^{2}}{\left(2m_{e}T_{e}\right)^{3/2}}\int_{0}^{\infty}\mathrm{e}^{-q^{2}/8m_{e}T_{e}}\left|\tilde{V}(q)\right|^{2} S_{ii}(q)q^{3}\,\mathrm{d}q \tag{19}$$

Equation (19) is a basis for various classical and semiclassical models of the conductivity of plasmas and gases, most notably those described in refs.[11], [12], [13].

## 4  THE LENARD-BALESCU CONDUCTIVITY MODEL

The conductivity model described above applies to a Lorentzian plasma, in which the electrons are scattered by a static potential (stationary ions in a static screening approximation). The latter restrictions may be lifted by replacing the Landau collision integral [13], equations (166) - (170), with the Lenard-Balescu collision integral [7],[14],[15] which, for $m_e \ll m_i$, takes the form

$$q_\mathbf{k} \nu(\mathbf{k}) \to -2\pi n_i \int q_{\mathbf{k+q}} \frac{\mathbf{q}\cdot\mathbf{k}}{k^2} \frac{d^3\mathbf{q}}{(2\pi)^3} \int_{-\infty}^{\infty} |V(\mathbf{q},\omega)|^2 \delta(\varepsilon_{\mathbf{k+q}} - \varepsilon_\mathbf{k} - \omega) \langle \delta(E_\mathbf{p} - E_{\mathbf{p-q}} - \omega) \rangle_\mathbf{p} \, d\omega \qquad (20)$$

which represents the collision frequency in terms of processes in which momentum $\mathbf{q}$ and energy $\omega$ are exchanged between an electron and an ion during a collision. In equation (20), $V(\mathbf{q},\omega) = V(\mathbf{q})/\varepsilon(\mathbf{q},\omega)$ is the dynamically screened potential and the notation $\langle \ \rangle_\mathbf{p}$ denotes the thermal average over the ion momenta, $\mathbf{p}$, which is generally assumed to be characterised by a Maxwell-Boltzmann distribution. Equation (20) reduces to the Landau collision integral if $|V(\mathbf{q},\omega)|^2 \delta(\varepsilon_{\mathbf{k+q}} - \varepsilon_\mathbf{k} - \omega)$ is replaced by $|V(\mathbf{q},0)|^2 \delta(\varepsilon_{\mathbf{k+q}} - \varepsilon_\mathbf{k}) = |\tilde{V}(\mathbf{q})|^2 \delta(\varepsilon_{\mathbf{k+q}} - \varepsilon_\mathbf{k})$, which then describes elastic collisions in a static potential. The nature of this approximation is examined later. However it should be noted, at this stage, that the dynamical screening involves the total dielectric function of the plasma, and thus, at the outset, treats all plasma components as a single system and on an equal footing.

In (20), the argument of the first $\delta$-function gives $\mathbf{q}\cdot\mathbf{k} = m_e\omega - \tfrac{1}{2}q^2$. Now, in the context of the integration over $\mathbf{q}$, the integrand, apart from the $\mathbf{q}\cdot\mathbf{k}$ factor, leads to an even function of $\omega$. The term $m_e\omega$ in the expression for $\mathbf{q}\cdot\mathbf{k}$ can therefore be dropped, since it gives rise to an integrand that is an odd function of $\omega$ and whose integral therefore vanishes. Equation (20) then becomes

$$q_\mathbf{k}\nu(\mathbf{k}) = \frac{\pi n_i}{k^2} \int q_{\mathbf{k+q}} q^2 \frac{d^3\mathbf{q}}{(2\pi)^3} \int_{-\infty}^{\infty} |V(\mathbf{q},\omega)|^2 \delta(\varepsilon_{\mathbf{k+q}} - \varepsilon_\mathbf{k} - \omega) \langle \delta(E_\mathbf{p} - E_{\mathbf{p-q}} - \omega) \rangle_\mathbf{p} \, d\omega \qquad (21)$$

Now, the resistivity (7) depends on

$$\langle\langle \varepsilon_\mathbf{k} \nu(\mathbf{k}) \rangle\rangle = \frac{T_B}{T_e} \frac{\pi n_i}{2m_e} \int q^2 \frac{d^3\mathbf{q}}{(2\pi)^3} \int_{-\infty}^{\infty} |V(\mathbf{q},\omega)|^2 \langle q_{\mathbf{k+q}} \delta(\varepsilon_{\mathbf{k+q}} - \varepsilon_\mathbf{k} - \omega) \rangle_\mathbf{k} \langle \delta(E_\mathbf{p} - E_{\mathbf{p-q}} - \omega) \rangle_\mathbf{p} \, d\omega \qquad (22)$$

which treats inelastic collisions in the dynamically screened potential, and moreover exhibits the required symmetry between the particles for treating binary collisions between electrons and ions.

In (22), the quantities $\langle q_{k+q}\delta(\varepsilon_{k+q}-\varepsilon_k-\omega)\rangle_k$ and $\langle \delta(E_p - E_{p-q} - \omega)\rangle_p$ are just the free-particle dynamic structure factors $S^0_{ee}(\mathbf{q},-\omega)$ and $S^0_{ii}(\mathbf{q},\omega)$, for the electrons and ions respectively. Hence (22) becomes

$$\langle\langle \varepsilon_k v(\mathbf{k})\rangle\rangle = \frac{T_B}{T_e}\frac{\pi n_i}{2m_e}\int q^2 \frac{d^3\mathbf{q}}{(2\pi)^3}\int_{-\infty}^{\infty}|V(\mathbf{q},\omega)|^2 S^0_{ee}(\mathbf{q},-\omega)S^0_{ii}(\mathbf{q},\omega)\,d\omega \qquad (23)$$

The non-interacting fermion dynamic structure factor, $S^0_{ee}$, is given exactly by the formula [7],[16],[17],

$$S^0_{ee}(\mathbf{q},\omega) = \langle q_{k+q}\delta(\varepsilon_k - \varepsilon_{k-q} - \omega)\rangle_k = \frac{1}{4uT_e}\left(\frac{1}{v}\right)^{1/2}\frac{L(u,v;\eta)}{I_{1/2}(\eta)} \qquad (24)$$

in which, $u = \omega/T_e$, $v = q^2/2m_e T_e$, where $T_e$ is the electron temperature, and

$$L(u,v;\eta) = \ln\left(\frac{1+\exp(\eta-(u-v)^2/4v)}{1+\exp(\eta-(u+v)^2/4v)}\right),$$ while the dynamic structure factor, $S^0_{ii}$ for non-

interacting Maxwell Boltzmann particles, which corresponds formally to the $\eta \to -\infty$ limit of (24), is [2], [18]

$$S^0_{ii}(\mathbf{q},\omega) = \langle\delta(E_p - E_{p-q} - \omega)\rangle_p = \sqrt{\frac{1}{4\pi E_q T_i}}\exp\left(-\frac{1}{4E_q T_i}(\omega - E_q)^2\right) \qquad (25)$$

where $T_i$ is the ion temperature. In (24), $I_j(\eta) = \int_0^{\infty}\frac{y^j}{1+\exp(y-\eta)}dy$ denotes a Fermi integral, which, for $j = \frac{1}{2}$, for a free electron gas, is given in terms of the temperature and density by

$$I_{1/2}(\eta) = 2\pi^2 n_e (2m_e T_e)^{-3/2} \qquad (26)$$

and, in terms of which [1],

$$T_B/T_e = I_{1/2}(\eta)/I'_{1/2}(\eta) \qquad (27)$$

### 4.1 The Lenard-Balescu conductivity in the Maxwell-Boltzmann limit

Let us now calculate the Lenard-Balescu resistivity for a Boltzmann distribution of electrons. This yields, from (22), for $T_e = T_i = T$,

$$\langle\langle\varepsilon_k v(\mathbf{k})\rangle\rangle = \frac{n_i}{4\pi m_e} \int q^2 \frac{d^3 \mathbf{q}}{4\pi} \int_{-\infty}^{\infty} |V(\mathbf{q},\omega)|^2 \langle\delta(\varepsilon_{\mathbf{k}+\mathbf{q}} - \varepsilon_\mathbf{k} - \omega)\rangle_\mathbf{k} \langle\delta(E_\mathbf{p} - E_{\mathbf{p}-\mathbf{q}} - \omega)\rangle_\mathbf{p} d\omega \quad (28)$$

$$= \frac{n_i}{4\pi m_e} \mathcal{L}_2^0(T)$$

where $\mathcal{L}_\nu^\mu(T)$ is defined by (209). Applying the result (223) from APPENDIX A yields

$$\mathcal{L}_2^0(T) = \sqrt{\frac{m}{2\pi T}} \int_0^\infty q^3 |\tilde{V}(q)|^2 S_{ii}(q) e^{-q^2/8mT} dq \quad (29)$$

in which $m$ is the reduced mass of the electron–ion system, and hence,

$$\langle\langle\varepsilon_k v(\mathbf{k})\rangle\rangle = \frac{n_i}{8\pi^2 m_e} \sqrt{\frac{2\pi m}{T}} \int_0^\infty q^3 |\tilde{V}(q)|^2 S_{ii}(q) e^{-q^2/8mT} dq \quad (30)$$

The various other factors in the formula (8) for the mean collision frequency may be evaluated, for the case of a Boltzmann distribution, using (178)-(179), as follows:

$$\langle\langle\varepsilon^{5/2}\rangle\rangle\langle\langle\varepsilon^{-1/2}\rangle\rangle = \frac{3}{2} \frac{T_B^2}{\pi^2 n_e} \frac{\langle k^3 \rangle}{1+e^{-\eta}} \xrightarrow{MB} \frac{24}{\pi} T^2 \quad (31)$$

and the mean collision frequency is then given by

$$\bar{\nu} = \frac{3T}{2} \frac{\langle\langle\varepsilon v \rangle\rangle}{\langle\langle\varepsilon^{5/2}\rangle\rangle\langle\langle\varepsilon^{-1/2}\rangle\rangle} = \frac{1}{32\sqrt{\pi}} \frac{m^2 n_i}{(2mT)^{3/2} m_e} \int_0^\infty q^3 |\tilde{V}(q)|^2 S_{ii}(q) e^{-q^2/8mT} dq \quad (32)$$

which yields the resistivity as

$$\sigma^{-1} = \frac{m_e \bar{\nu}}{n_e e^2} = \frac{1}{32\sqrt{\pi} Z e^2} \frac{m^2}{(2mT)^{3/2}} \int_0^\infty q^3 |\tilde{V}(q)|^2 S_{ii}(q) e^{-q^2/8mT} dq \quad (33)$$

which, apart from the reduced mass $m$ of the electron-ion system appearing everywhere in place of the electron mass, $m_e$, agrees exactly with (19).

### 4.2 The general Lenard-Balescu result

To help understand this better, let us return to the general expression for $\langle\langle\varepsilon_k v(\mathbf{k})\rangle\rangle$ given by (23), which we write in the form

$$\langle\langle\varepsilon_k v(\mathbf{k})\rangle\rangle = \frac{T_B}{T} \frac{\pi n_i}{2m_e} \int q^2 \frac{d^3 \mathbf{q}}{(2\pi)^3} \int_{-\infty}^{\infty} |V(\mathbf{q})|^2 \frac{1}{|\varepsilon(\mathbf{q},\omega)|^2} S_{ee}^0(\mathbf{q},-\omega) S_{ii}^0(\mathbf{q},\omega) d\omega \quad (34)$$

and define the effective ion dielectric function by

$$\tilde{\varepsilon}_i(\mathbf{q},\omega) = \frac{\varepsilon(\mathbf{q},\omega)}{\varepsilon_e(\mathbf{q},\omega)} \qquad (35)$$

where $\varepsilon_e(\mathbf{q},\omega)$ is the dielectric function due to the electrons alone. The function $\tilde{\varepsilon}_i(\mathbf{q},\omega)$ is the ion dielectric function *modified by the presence of the electrons*. (This is the significance of the ~ notation.) It should however be understood that the rationale for this factorisation is ultimately dependent upon $m_i \gg m_e$. In the Random Phase or Ring approximation, the corresponding interacting structure factors are given by the dielectric superposition principle [4]

$$S_{ee}(\mathbf{q},\omega) = \frac{S_{ee}^0(\mathbf{q},\omega)}{\left|\varepsilon_e(\mathbf{q},\omega)\right|^2}$$

$$\tilde{S}_{ii}(\mathbf{q},\omega) = \frac{S_{ii}^0(\mathbf{q},\omega)}{\left|\tilde{\varepsilon}_i(\mathbf{q},\omega)\right|^2} \qquad (36)$$

whereupon (34) becomes

$$\langle\langle \varepsilon_\mathbf{k} \nu(\mathbf{k}) \rangle\rangle = \frac{T_B}{T} \frac{\pi n_i}{2m_e} \int q^2 \frac{d^3\mathbf{q}}{(2\pi)^3} \int_{-\infty}^{\infty} |V(\mathbf{q})|^2 S_{ee}(\mathbf{q},-\omega) \tilde{S}_{ii}(\mathbf{q},\omega) d\omega \qquad (37)$$

so that the treatment of the dynamical screening has been transferred from the potential, which now appears as the unscreened potential, to the structure factors.

The mean collision frequency, and resistivity, calculated from (37), making use of (8), (10) and (7), are therefore

$$\bar{\nu} = \frac{\pi n_e n_i}{4m_e \langle k^3 \rangle} \frac{1+e^{-\eta}}{T} \int q^2 |V(\mathbf{q})|^2 \frac{d^3\mathbf{q}}{4\pi} \int_{-\infty}^{\infty} S_{ee}(\mathbf{q},-\omega) \tilde{S}_{ii}(\mathbf{q},\omega) d\omega \qquad (38)$$

$$\sigma^{-1} = \frac{\pi n_i}{4e^2 \langle k^3 \rangle} \frac{1+e^{-\eta}}{T} \int q^2 |V(\mathbf{q})|^2 \frac{d^3\mathbf{q}}{4\pi} \int_{-\infty}^{\infty} S_{ee}(\mathbf{q},-\omega) \tilde{S}_{ii}(\mathbf{q},\omega) d\omega \qquad (39)$$

### 4.3 The generalized Ziman conductivity formula

For electron-ion scattering, we have $m_e \ll m_i$, and hence that $\tilde{S}_{ii}(\mathbf{q},\omega)$ is peaked and likely to have most of its strength in the small-$\omega$ domain of $S_{ee}(\mathbf{q},\omega)$, conditions for which are $\Omega_i \ll \Omega_e, T$. The following approximation can then be made:

$$\int_{-\infty}^{\infty} S_{ee}(\mathbf{q},-\omega) \tilde{S}_{ii}(\mathbf{q},\omega) \, d\omega \simeq \lim_{\omega \to 0}\left(S_{ee}(\mathbf{q},-\omega)\right) \int_{-\infty}^{\infty} \tilde{S}_{ii}(\mathbf{q},\omega) \, d\omega$$

$$= \lim_{\omega \to 0}\left(S_{ee}(\mathbf{q},-\omega)\right) \tilde{S}_{ii}(\mathbf{q}) \quad (40)$$

where $\tilde{S}_{ii}(\mathbf{q})$ is the ion static structure factor. This approximation underlies, and is a generalization of, the result (221) due to Boercker et al [19]. It is examined in more detail in section 8.

Now,

$$\lim_{\omega \to 0}\left(S_{ee}(\mathbf{q},-\omega)\right) = \frac{1}{\left|\varepsilon_e(q,0)\right|^2} \lim_{\omega \to 0}\left(S_{ee}^0(\mathbf{q},-\omega)\right) \quad (41)$$

and, using the exact analytical formula for the non-interacting fermion structure factor (24) along with (26), the zero-frequency limit of the electron structure factor can be determined as follows:

$$\lim_{\omega \to 0}\left(S_{ee}^0(\mathbf{q},-\omega)\right) = \frac{1}{4q}\left(\frac{2m_e}{T}\right)^{1/2} \frac{1}{I_{1/2}(\eta)} \lim_{u \to 0}\left(\frac{L(-u,v;\eta)}{-u}\right)$$

$$= \frac{1}{4q}\left(\frac{2m_e}{T}\right)^{1/2} \frac{1}{I_{1/2}(\eta)} \frac{1}{1+\exp(q^2/8m_e T - \eta)} \quad (42)$$

$$= \frac{1}{4q}\left(\frac{2m_e}{T}\right)^{1/2} \frac{1}{I_{1/2}(\eta)} p(\tfrac{1}{2}q)$$

$$= \frac{m_e^2 T}{2\pi^2 n_e q} p(\tfrac{1}{2}q)$$

Hence, making the approximation (40), and the substitutions (41) - (42), equations (38) and (39) reduce to

$$\bar{\nu} = \frac{n_i m_e}{8\pi} \frac{1+e^{-\eta}}{\langle k^3 \rangle} \int_0^{\infty} p(\tfrac{1}{2}q) \left|\tilde{V}(q)\right|^2 \tilde{S}_{ii}(q) q^3 \, dq \quad (43)$$

$$\sigma^{-1} = \frac{m_e^2}{8\pi Z e^2} \frac{1+e^{-\eta}}{\langle k^3 \rangle} \int_0^{\infty} p(\tfrac{1}{2}q) \left|\tilde{V}(q)\right|^2 \tilde{S}_{ii}(q) q^3 \, dq \quad (44)$$

which agree with (15) and (16) respectively, except that, in (43) - (44), the ion structure factor is provided explicitly by

$$\tilde{S}_{ii}(q) = \int_{-\infty}^{\infty} S_{ii}^0(\mathbf{q},\omega) \left| \tilde{\varepsilon}_i(\mathbf{q},\omega) \right|^{-2} d\omega \qquad (45)$$

$$= \int_{-\infty}^{\infty} S_{ii}^0(\mathbf{q},\omega) \left| \frac{\varepsilon_e(\mathbf{q},\omega)}{\varepsilon(\mathbf{q},\omega)} \right|^2 d\omega$$

where $S_{ii}^0(\mathbf{q},\omega)$ is the non-interacting structure factor for the ions alone, as given by (25). Furthermore, the statically screened electron-ion potential appearing in (43) - (44) is given by

$$\tilde{V}(q) = \frac{V(q)}{\varepsilon_e(q,0)} \qquad (46)$$

which involves screening by electrons only, thus confirming the earlier assertion. This asymmetry can now be seen to be a consequence of the electrons having very small masses compared with the ions.

## 5 DISCUSSION

The following observations can now be made.

The static approximation embodied in the Landau collision integral, and in the standard Ziman formula and its generalization (44), is a consequence of either the approximation (40), or, equivalently, in the Boltzmann limit, that of Boercker *et al* as given by (221), which depend upon $m_e \ll m_i$,

The derivation from the dynamic (Lenard-Balescu) collision integral, which allows energy transfer between electrons and ions, replaces this structure factor with the complete static ion structure factor (as per the Ziman formula) but modified by the presence of the electrons in accordance with (45). Otherwise the formulae are unchanged.

Ion dynamical modes (plasma modes) are therefore implicitly included in this formalism through the integral over $\omega$. However, as the plasma modes correspond to zeros of $\tilde{\varepsilon}_i(q,\omega)$, which are singularities of $\tilde{S}_{ii}(q,\omega)$, due account needs to be taken of them. This done in section 8.2.

A modal decomposition of the ion structure factor into its ground-state and dynamic (collective) parts is provided by equations (162-165) of ref. [17]:

$$\tilde{S}_{ii}(\mathbf{q},\omega) = F_0(\mathbf{q})\delta_\mathbf{q}^i(\omega) + \left(1 - F_0(\mathbf{q})\right) \frac{q^2}{2m_i} g_\mathbf{q}^i(\omega) \qquad (47)$$

where

$$\int_{-\infty}^{\infty} \delta_{\mathbf{q}}(\omega) \, d\omega = 1 \tag{48}$$

$$\int_{-\infty}^{\infty} \delta_{\mathbf{q}}(\omega) \omega \, d\omega = \frac{q^2}{2m_i} \tag{49}$$

$$\int_{-\infty}^{\infty} g_{\mathbf{q}}(\omega) \, d\omega = \frac{1}{\Omega_{\mathbf{q}}} \coth\left(\frac{\Omega_{\mathbf{q}}}{2T}\right) \tag{50}$$

$$\int_{-\infty}^{\infty} g_{\mathbf{q}}(\omega) \omega \, d\omega = 1 \tag{51}$$

$$F_0(\mathbf{q}) = 1 - \left(\frac{\left(\Omega_{\mathbf{q}}/\Omega_i\right)^2}{\left(\Omega_{\mathbf{q}}/\Omega_i\right)^2 - \lambda q^2 D_e^2}\right)\left(\frac{1}{1 + \lambda\left(1 + q^2 D_e^2\right)}\right) \tag{52}$$

$\Omega_{\mathbf{q}}$ is the frequency of the ion plasma mode with wavevector $\mathbf{q}$, $\lambda = D_i^2/D_e^2 = T/ZT_B$, and $D_e$ and $D_i$ are the plasma and ion effective screening lengths respectively (see below). (Note that this gives the particle density (as opposed to charge density) structure factor, and is normalised by the ion, rather than the electron density, so the factor of $w_i = Z^2 n_i/n_e$, included in ref [17], does not appear here.)

In the weak damping limit, $g_{\mathbf{q}}^i(\omega)$ is given by, [17]

$$g_{\mathbf{q}}^i(\omega) = \frac{1}{\Omega_{\mathbf{q}}}\left(\left(\mathcal{N}_{\mathbf{q}}+1\right)\delta(\omega - \Omega_{\mathbf{q}}) + \mathcal{N}_{\mathbf{q}}\delta(\omega + \Omega_{\mathbf{q}})\right) \tag{53}$$

where $\mathcal{N}_{\mathbf{q}}$ is the Bose-Einstein function,

$$\mathcal{N}_{\mathbf{q}} = \frac{1}{\exp(\Omega_{\mathbf{q}}/T) - 1} \tag{54}$$

which gives the equilibrium excitation of plasma modes. The two terms in the expression (53) correspond to Brillouin processes in which a collective excitation mode is created or destroyed in a scattering event.

In the first instance, the screening lengths $D_e$, $D_i$ and $D$ are defined by

$$\frac{1}{D_i^2} = q^2\left(\varepsilon_i(\mathbf{q},0) - 1\right)$$

$$\frac{1}{D_e^2} = q^2\left(\varepsilon_e(\mathbf{q},0) - 1\right) \tag{55}$$

$$\frac{1}{D^2} = q^2 \left( \varepsilon(\mathbf{q}, \omega) - 1 \right) \simeq \frac{1}{D_e^2} + \frac{1}{D_i^2} \tag{56}$$

which define these quantities as being, in general, functions of $q$. However, for $q$ not too large (ie in the classical or semiclassical regimes) then $D_e$, $D_i$ and $D$ can be replaced by their $\mathbf{q} = 0$ limits, which corresponds, in the RPA, to the Debye approximation and yields the Debye lengths with the degeneracy correction, represented by the factor $R_0 = I'_{1/2}(\eta)/I_{1/2}(\eta) = T/T_B$ [1], [17][a], [20], [21], already built-in. The appropriate generalizations, in the form of Padé approximants, to large values of $q$ can be found in [17]. Here, it is sufficient to note that, for $q \to \infty$, $D^2 \to \mathcal{O}(q^2)$.

Equation (165) in [17], with $\gamma_i = 3$, implies the following expression for $(\Omega_\mathbf{q}/\Omega_i)^2$ which incorporates the $q$-dependence of the intrinsic modes.

$$\left( \frac{\Omega_\mathbf{q}}{\Omega_i} \right)^2 = \frac{q^2 D_e^2}{1 + q^2 D_e^2} + q^2 D_i^2 \left( \gamma_i + \frac{D_i^2}{D_e^2} \frac{\gamma_i - 1}{(1 + 2q^2 D_i^2)(\gamma_i - 2 + q^2 D_i^2)} \right)$$

$$= \frac{q^2 D_e^2}{1 + q^2 D_e^2} + q^2 D_i^2 \left( 3 + \frac{D_i^2}{D_e^2} \frac{2}{(1 + 2q^2 D_i^2)(1 + q^2 D_i^2)} \right) \tag{57}$$

The static structure factor that follows from (47) is

$$\tilde{S}_{ii}(q) \equiv \int_{-\infty}^{\infty} \tilde{S}_{ii}(\mathbf{q}, \omega) \, d\omega = F_0(q) + \left( 1 - F_0(q) \right) \frac{q^2}{2 m_i \Omega_\mathbf{q}} \coth\left( \frac{\Omega_\mathbf{q}}{2T} \right) \tag{58}$$

which is a modal decomposition of the static structure factor into independent particle and collective parts.

An alternative decomposition of the ion static structure factor (58) is provided by

$$\tilde{S}_{ii}(q) = S_0(q) + \left(1 - S_0(q)\right)(qD_i)^2 \left[ \left(\frac{\Omega_\mathbf{q}}{\Omega_i}\right)^2 - (qD_i)^2 \right]^{-1} \left( \frac{\Omega_\mathbf{q}}{2T} \coth\left(\frac{\Omega_\mathbf{q}}{2T}\right) - 1 \right) \tag{59}$$

where

---

[a] Note that, in [17], $D_e$ is defined to be the classical Debye length without the factor $R_0$.

$$S_0(q) = F_0(\mathbf{q}) + (1 - F_0(\mathbf{q}))\frac{q^2 T}{m_i \Omega_{\mathbf{q}}^2}$$

$$= 1 - \frac{1/q^2 D_i^2}{1 + 1/q^2 D^2} \quad (60)$$

$$= 1 - \frac{\varepsilon_i(\mathbf{q},0) - 1}{\varepsilon(\mathbf{q},0)}$$

is the (semi)classical[b] ion static structure factor [17], [22]. Equation (59) corresponds to a classical-quantal decomposition of the static structure factor [17]. The classical limit corresponds to high temperatures, $T \gg \Omega_{\mathbf{q}}$, when the second term is $\mathcal{O}\left((\Omega_{\mathbf{q}}/T)^2\right)$. Equation (45) then implies that, in the classical limit,

$$S_0(q) \equiv \frac{\varepsilon_e(\mathbf{q},0)}{\varepsilon(\mathbf{q},0)} = \int_{-\infty}^{+\infty} S_{ii}^0(\mathbf{q},\omega)\left|\frac{\varepsilon_e(\mathbf{q},\omega)}{\varepsilon(\mathbf{q},\omega)}\right|^2 d\omega \quad (61)$$

Note that equation (61) is a property of the classical limit of the RPA and does not depend upon $m_e \ll m_i$. It should also hold in the regimes of arbitrary electron degeneracy (provided that $T \gg \Omega_{\mathbf{q}}$).

Substituting the classical limit of the non-interacting dynamical structure factor given by (25), for the ions, yields

$$S_0(q) \equiv \frac{\varepsilon_e(\mathbf{q},0)}{\varepsilon(\mathbf{q},0)} = \frac{1}{\sqrt{2\pi}}\frac{1}{qv}\int_{-\infty}^{+\infty}\exp\left(-\frac{\omega^2}{2q^2 v^2}\right)\left|\frac{\varepsilon_e(\mathbf{q},\omega)}{\varepsilon(\mathbf{q},\omega)}\right|^2 d\omega \quad (62)$$

where $v^2 = T/m_i$. Equation (61) is thus identified as a generalization of (218) to which it reduces when $m_e \ll m_i$.

## 6   GENERALIZATION TO LOW TEMPERATURES

The approximation (40) does not extend to arbitrarily low temperatures. This is reflected in the non-commuting nature of the $T \to 0$ and $\omega \to 0$ limits. In order to achieve the correct low-temperature behaviour, the following formula for $S_{ee}^0(\mathbf{q},\omega)$ must be substituted for the $\omega \to 0$ limit, (42),

---

[b] *Semi*classical, in the sense that electron degeneracy is accounted for.

$$S_{ee}^0(\mathbf{q},\omega) \simeq \frac{1}{4q}\left(\frac{2m_e}{T}\right)^{1/2} \frac{p(\tfrac{1}{2}q)}{I_{1/2}(\eta)} \frac{\omega/T}{1-\exp(-\omega/T)}$$

$$= \frac{m_e^2 T}{2\pi^2 n_e q} p(\tfrac{1}{2}q) \frac{\omega/T}{1-\exp(-\omega/T)}$$

(63)

Equation (63) represents an approximation to (24) that holds in the degenerate limit for all $\omega/T$, and for $\omega \ll T$. It also satisfies, for all $\omega/T$, the Kubo-Martin-Schwinger (KMS) detailed-balance relation [7]

$$S(\mathbf{q},-\omega) = \exp(-\omega/T) S(\mathbf{q},\omega) \qquad (64)$$

which is a fundamental property of any equilibrium dynamic structure factor. Equation (63) contains the correct limits for both $T \to 0$ and $\omega \to 0$. Substituting (63) into (38) yields

$$\bar{\nu} = \frac{n_i m_e}{8\pi} \frac{1+e^{-\eta}}{\langle k^3 \rangle} \int_0^\infty p(\tfrac{1}{2}q)|\tilde{V}(\mathbf{q})|^2 q^3 \, dq \int_{-\infty}^\infty \frac{\omega/T}{\exp(\omega/T)-1} \tilde{S}_{ii}(\mathbf{q},\omega) \, d\omega \qquad (65)$$

in place of (43). Equation (65) represents a further generalization of the Ziman formula, one which extends (43) to arbitrarily low temperatures in regimes of high degeneracy.

Substituting the collective ion Structure factor, $\tilde{S}_{ii}(\mathbf{q},\omega) = (1-F_0(q))\frac{q^2}{2m_i} g_\mathbf{q}^i(\omega)$ from (47), with $g_\mathbf{q}^i(\omega)$ given by (53), yields

$$\bar{\nu} = \frac{n_i m_e}{8m_i\pi} \frac{1+e^{-\eta}}{\langle k^3 \rangle T} \int_0^\infty (1-F_0(q)) p(\tfrac{1}{2}q)|\tilde{V}(q)|^2 \frac{\exp(\Omega_\mathbf{q}/T)}{(\exp(\Omega_\mathbf{q}/T)-1)^2} q^5 \, dq \qquad (66)$$

The Ziman formula is generally considered to be applicable in the liquid metal regime, ie in degenerate regimes at low temperatures, possibly extending below the ion plasma frequency, $\Omega_i$. At such low temperatures, the conditions for ideality are unlikely to be satisfied by the ions, and use of formulae based purely upon the RPA become inapplicable. Short range correlations will modify the structure factor and will, generally impose an upper limit on the wavenumbers contributing to the integrals (93) - (98) analogous to that in the Debye model of a crystalline solid.

However, in a disordered system, the upper limit is less well defined than in a crystalline solid. One approach is to *impose* a cutoff $K$ on the structure factor $S_0(q)$ so that $1-S_0(q)$ vanishes for $q > K$, where $K$ is defined by

$$\int_0^K (1-S_0(q)) q^2 \, dq = 2\pi^2 n_i \qquad (67)$$

If $S_0(q)$ in (67) is taken to be the classical structure factor as given by the Debye-Hückel approximation (60) (in which $D_e$, $D_i$ etc are given by their values at $q=0$) the cutoff is, for weakly coupled plasmas, yielded as $\pi/2L$, where $L = Z^2 e^2/4\pi\varepsilon_0 T$ is the Landau length; and as the Debye radius, $(6\pi^2 n_i)^{1/3}$, for strongly coupled ones. However, for strongly-coupled plasmas, the shape of $S_0(q)$ for $q < K$ is unlikely to be correct and the discontinuity at $q = K$ can have manifest consequences. If the short-range correlations are correctly included in $S_0(q)$, then the following should hold [3]

$$\int_0^\infty (1 - S_0(q)) q^2 \, dq = 2\pi^2 n_i \qquad (68)$$

without the need for an artificially imposed cutoff, it being the condition that the ion-ion pair-distribution function vanishes at zero separation. While the integral (68) diverges in the Debye-Hückel approximation due to the absence of short-range correlations, it is actually convergent in the RPA when the large-$q$ quantum corrections are included. However, the RPA does not model short-range correlations, so there is no reason to suppose that (68) will automatically be satisfied. Even in the case of a structure factor satisfying (68), it will generally be possible to identify a value $q = K$ beyond which $S_0(q) \simeq 1$, and which thereby approximately satisfies (67).

Dense plasmas, in the low temperature regime, are considered to be strongly-coupled ($\Gamma \equiv Z^2 e^2 (4\pi n_i/3)^{1/3}/4\pi\varepsilon_0 T \gg 1$) and strongly degenerate ($\eta \gg 1 \Rightarrow \lambda = 3T/2ZT_F \ll 1$). The appropriate cut-off for the structure factor is taken to be the Debye radius, $K = (6\pi^2 n_i)^{1/3} = 2k_F/(4Z)^{1/3}$. Therefore, provided that $4Z > 1$ it follows that $K < 2k_F$, and the integral (66) can then be written,

$$\overline{V} = \frac{n_i m_e}{4 m_i \pi} \frac{1}{k_F^3 T} \int_0^K (1 - F_0(q)) |\tilde{V}(q)|^2 \frac{\exp(\Omega_q/T)}{(\exp(\Omega_q/T) - 1)^2} q^5 \, dq \qquad (69)$$

Therefore, referring to (52), since $\lambda \ll 1$, $1 - F_0(q) \simeq 1$,

$$\bar{\nu} = \frac{\langle \nu_0 k^3 \rangle}{3\pi^2 n_e} \frac{1}{m_i T} \int_0^K \left( \frac{q^2 D_e^2}{1 + q^2 D_e^2} \right)^2 \frac{\exp(\Omega_q/T)}{(\exp(\Omega_q/T) - 1)^2} q \, dq$$

$$= \frac{\langle \nu_0 k^3 \rangle}{3\pi^2 n_e} \frac{1}{m_i D_e^2 T} \int_0^K \left( \frac{q^2 D_e^2}{1 + q^2 D_e^2} \right)^3 \frac{\exp(\Omega_q/T)}{(\exp(\Omega_q/T) - 1)^2} (1 + q^2 D_e^2) \frac{dq}{q} \qquad (70)$$

$$= \lambda \frac{\langle \nu_0 k^3 \rangle}{3\pi^2 n_e} \left( \frac{\Omega_i}{T} \right)^2 \int_0^K \left( \frac{\Omega_q}{\Omega_i} \right)^6 \frac{\exp(\Omega_q/T)}{(\exp(\Omega_q/T) - 1)^2} (1 + q^2 D_e^2) \frac{dq}{q}$$

in which (neglecting terms of $\mathcal{O}(\lambda)$)

$$\left( \frac{\Omega_q}{\Omega_i} \right)^2 = \frac{q^2 D_e^2}{1 + q^2 D_e^2} \qquad (71)$$

and where $\nu_0(k) \propto k^{-3}$ is the classical collision frequency, (80).

The integral (70) can be transformed to one over $u = \Omega_q/T$, by means of

$$\frac{1}{1 + q^2 D_e^2} \frac{dq}{q} = \frac{du}{u} \qquad (72)$$

which follows from (71). Hence

$$\bar{\nu} = \lambda \frac{\langle \nu_0 k^3 \rangle}{3\pi^2 n_e} \left( \frac{T}{\Omega_i} \right)^4 \int_0^{\Theta/T} u^5 \frac{e^u}{(e^u - 1)^2} \left[ 1 - \left( \frac{uT}{\Omega_i} \right)^2 \right]^{-2} du$$

$$= \frac{\langle \nu_0 k^3 \rangle}{2\pi^2 n_e} \frac{\Omega_i}{ZT_F} \left( \frac{T}{\Omega_i} \right)^5 \int_0^{\Theta/T} u^5 \frac{e^u}{(e^u - 1)^2} \left[ 1 - \left( \frac{uT}{\Omega_i} \right)^2 \right]^{-2} du \qquad (73)$$

where

$$\Theta = \Omega_K = \Omega_i \sqrt{\frac{K^2 D_e^2}{1 + K^2 D_e^2}} \qquad (74)$$

But $K^2 D_e^2 = \frac{16}{3}(4Z)^{-2/3}(T_F/\Omega_e)^2$, which is typically sufficiently small compared with unity, in plasmas and metals at densities $\lesssim$ few × solid density, for

$$\left(\frac{uT}{\Omega_i}\right)^2 \le \left(\frac{\Theta}{\Omega_i}\right)^2 = \frac{K^2 D_e^2}{1+K^2 D_e^2} \tag{75}$$

to be neglected in (73). This yields

$$\bar{\nu} = \frac{\langle V_0 k^3 \rangle}{2\pi^2 n_e} \frac{\Omega_i}{ZT_F} \left(\frac{T}{\Omega_i}\right)^5 \int_0^{\Theta/T} u^5 \frac{e^u}{(e^u-1)^2} du \tag{76}$$

Equation (76) is just the famous Bloch formula [23], [24] for the effective collision frequency. The resistivity which follows from (76) is

$$\sigma^{-1} = \frac{\langle V_0 k^3 \rangle}{2\pi^2 \epsilon_0 n_e \Omega_e^2} \frac{\Omega_i}{ZT_F} \left(\frac{T}{\Omega_i}\right)^5 \int_0^{\Theta/T} u^5 \frac{e^u}{(e^u-1)^2} du \tag{77}$$

In application to metallic solids, the collective modes are phonons, for which $\Theta = \Theta_R$, which corresponds approximately to the Debye temperature, $\Theta_D \simeq Kc$, where $c$ is the sound velocity. Equation (77) then yields the Gruneisen-Bloch formula for the resistivity of a normal metal [24]. In plasmas, the collective modes are the plasma ion-acoustic modes, for which $\Theta$ is given by (74). However, it is observed that the sound velocity in a liquid metal is not too different from $\Omega_i D_e$, so that $\Theta \approx \Theta_R \approx \Theta_D$ A notable property of equations (76) - (77) is that they exhibit the well-known $\sim T^5$ behaviour [24] as $T \to 0$.

A further observation is that equation (65) reduces to the Ziman form (43) only if $T \gg \Theta$.

The ion-acoustic Debye temperature, (74), increases with electron density as $n_e^{1/6}$, which means that, at very high densities, the factor $\left(1-(uT/\Omega_i)^2\right)^{-2}$ in (73) may deviate significantly from unity. By virtue of (75), $uT/\Omega_i < 1$, so one can use the binomial expansion to determine the corrections. This yields

$$\bar{\nu} = \frac{\langle V_0 k^3 \rangle}{2\pi^2 n_e} \frac{\Omega_i}{ZT_F} \sum_{n=1}^{\infty} n \left(\frac{T}{\Omega_i}\right)^{2n+3} \mathcal{J}_{2n+3}\left(\frac{\Theta}{T}\right) \tag{78}$$

which is now an exact representation of (70) - (71), and where $\mathcal{J}_n(z)$ denotes the Debye integral

$$\mathcal{J}_n(z) = \int_0^z u^n \frac{e^u}{(e^u-1)^2} du \tag{79}$$

## 7 COULOMB LOGARITHM

### 7.1 Coulomb logarithm in the generalized Ziman conductivity formula

The Coulomb Logarithm is defined in terms of the classical Coulomb collision frequency

$$\nu_0(k) = 4\pi n_i v a_\mathbf{k}^2 = \frac{n_i m_e}{4\pi} k |V(k)|^2 \tag{80}$$

through the conductivity-related standard collision frequency,

$$\nu_c = \overline{\nu_0} = \frac{1}{2} \frac{\langle \nu_0 k^3 \rangle}{\langle k^3 \rangle} = \frac{2\pi n_i m_e}{\langle k^3 \rangle} \left( \frac{Ze^2}{4\pi\varepsilon_0} \right)^2 \equiv \frac{\langle \nu_c k^3 \rangle}{\langle k^3 \rangle} \tag{81}$$

obtained by substituting $\nu_0(k)$ for $\nu$ in (8) and making use of (179), by

$$\ln \overline{\Lambda} = \frac{\overline{\nu}}{\overline{\nu_0}} = \frac{\overline{\nu}}{\nu_c} \tag{82}$$

With $\overline{\nu}$ given by (43), this yields

$$\ln \overline{\Lambda} = (1 + e^{-\eta}) \int_0^\infty p(\tfrac{1}{2}q) \tilde{S}_{ii}(q) \frac{1}{|\varepsilon_e(q,0)|^2} \frac{dq}{q}$$

$$= (1 + e^{-\eta}) \int_0^\infty p(\tfrac{1}{2}q) \tilde{S}_{ii}(q) \left( \frac{q^2 D_e^2}{1 + q^2 D_e^2} \right)^2 \frac{dq}{q} \tag{83}$$

In which $\varepsilon_e(q,0)$ has been taken to be given by

$$\varepsilon_e(q,0) = 1 + \frac{1}{q^2 D_e^2} \tag{84}$$

where $D_e$ is the effective screening length (modified, as appropriate, to take account of electron degeneracy).

Substituting the classical ion structure factor (60) into the formula (83) for the Coulomb logarithm yields

$$\ln \bar{\Lambda}_0 = \left(1+e^{-\eta}\right)\int_0^\infty p\left(\tfrac{1}{2}q\right) S_0(q) \frac{1}{\left|\varepsilon_e(q,0)\right|^2} \frac{dq}{q}$$

$$= \left(1+e^{-\eta}\right)\int_0^\infty p\left(\tfrac{1}{2}q\right) \frac{1}{\varepsilon(q,0)\varepsilon_e(q,0)} \frac{dq}{q} \tag{85}$$

$$= \frac{\lambda}{2}\left(1+e^{-\eta}\right)\int_0^\infty \frac{1}{1+\exp\left(x\Omega_e^2/8TT_B - \eta\right)} \frac{x}{(1+x)(1+\lambda+\lambda x)} dx$$

in which the integration variable has been transformed to $x = q^2 D_e^2$. Note that the integral is convergent at both limits, with the large-$q$ convergence being ensured by the essentially quantum-mechanical factor $p\left(\tfrac{1}{2}q\right)$, the cutoff being around the electron thermal deBroglie wavenumber, $m_e v$, for Maxwellian plasmas; or $\sim 2m_e v_F$ for degenerate plasmas. Equation (85) provides the conductivity Coulomb logarithm for hot Lorentzian plasmas, for which $T \gg \Omega_i$.

In order to generalize these formulae to include the low-temperature regime, equation (43) needs to be replaced by (65), for which the Coulomb logarithm is

$$\ln \bar{\Lambda} = \left(1+e^{-\eta}\right)\int_0^\infty p\left(\tfrac{1}{2}q\right) \frac{1}{\left|\varepsilon_e(q,0)\right|^2} \frac{dq}{q} \int_{-\infty}^\infty \frac{\omega/T}{\exp(\omega/T)-1} \tilde{S}_{ii}(\mathbf{q},\omega) d\omega \tag{86}$$

which replaces (83). In the low-temperature, near-solid-density regime, this yields, from (76),

$$\ln \bar{\Lambda}_{\text{Bloch}} = \frac{3}{2}\frac{\Omega_i}{ZT_F}\left(\frac{T}{\Omega_i}\right)^5 \int_0^{\Theta/T} u^5 \frac{e^u}{(e^u-1)^2} du \tag{87}$$

which is the Coulomb Logarithm corresponding to the Bloch resistivity (77).

### 7.2 Ion collective contribution

We now consider the ion collective contribution to the Coulomb logarithm in regimes not addressed in section 6. Returning to the 'exact' formula (38) and substituting the collective part of the ion structure factor, $\tilde{S}_{ii}(\mathbf{q},\omega) = \left(1-F_0(\mathbf{q})\right)\frac{q^2}{2m_i} g_\mathbf{q}^i(\omega)$ from (47), with $g_\mathbf{q}^i(\omega)$ given by (53), into (38). while making use of the KMS relation (64) for $S_{ee}(\mathbf{q},\omega)$, yields

$$\bar{\nu} = \frac{\pi n_e n_i}{4 m_e m_i \langle k^3 \rangle T}(1+e^{-\eta})\int \frac{1}{\Omega_q}(1-F_0(\mathbf{q}))q^4 |V(\mathbf{q})|^2 \frac{S_{ee}(\mathbf{q},\Omega_q)}{\exp(\Omega_q/T)-1}\frac{d^3\mathbf{q}}{4\pi} \quad (88)$$

For $m_e \ll m_i$, the random phase approximation gives

$$S_{ee}(\mathbf{q},\Omega_q) \simeq \frac{S_{ee}^0(\mathbf{q},\Omega_q)}{|\varepsilon_e(\mathbf{q},0)|^2} \quad (89)$$

where

$$S_{ee}^0(\mathbf{q},\Omega_q) = \frac{1}{2q}\sqrt{\frac{m_e}{2T}}\frac{1}{I_{1/2}(\eta)}\frac{1}{1-\exp(-\Omega_q/T)}\ln\left(\frac{1+e^{\eta-\Delta_q/T}}{1+e^{\eta-\Delta/T-\Omega_q/T}}\right)$$
$$= \frac{1}{q}\frac{m_e^2 T}{2\pi^2 n_e}\frac{1}{1-\exp(-\Omega_q/T)}\ln\left(\frac{1+e^{\eta-\Delta_q/T}}{1+e^{\eta-\Delta_q/T-\Omega_q/T}}\right) \quad (90)$$

is the dynamic structure factor for the ideal electron gas (24), evaluated at $\omega = \Omega_q$, and

$$\Delta_q = \frac{1}{4y}(1-y)^2$$
$$y = \frac{q^2}{2 m_e \Omega_q} \quad (91)$$

Substituting (90) into (88) leads to

$$\bar{\nu} = \frac{n_i m_e}{8\pi m_i \langle k^3 \rangle}(1+e^{-\eta})\int_0^\infty \frac{1}{\Omega_q}(1-F_0(q))|\tilde{V}(q)|^2 \frac{\exp(\Omega_q/T)}{(\exp(\Omega_q/T)-1)^2}\ln\left(\frac{1+e^{\eta-\Delta_q/T}}{1+e^{\eta-\Delta_q/T-\Omega_q/T}}\right) q^5 \, dq$$

$$= \nu_0(1+e^{-\eta})\int_0^\infty \frac{1}{m_i\Omega_q}\frac{1-F_0(q)}{|\varepsilon_e(q,0)|^2}\frac{\exp(\Omega_q/T)}{(\exp(\Omega_q/T)-1)^2}\ln\left(\frac{1+e^{\eta-\Delta_q/T}}{1+e^{\eta-\Delta_q/T-\Omega_q/T}}\right) q \, dq \quad (92)$$

where $\nu_0$ is the standard collision frequency given by (81). Changing the integration variable to $x = q^2 D_e^2$ yields

$$\bar{\nu} = \nu_0 \lambda \frac{\Omega_i}{2T}(1+e^{-\eta})\int_0^\infty \tilde{B}^i(x)\left(\frac{x}{1+x}\right)^{3/2}\frac{\exp(\Omega^i(x)/T)}{(\exp(\Omega^i(x)/T)-1)^2}\ln\left(\frac{1+e^{\eta-\Delta(x)/T}}{1+e^{\eta-\Delta(x)/T-\Omega^i(x)/T}}\right) dx \quad (93)$$

and where, referring to (52) and (57),

$$\tilde{B}^i(x) = \frac{\Omega_i}{\Omega^i(x)}\sqrt{\frac{x}{1+x}} B^i(x)$$

$$= B^i(x)\left[1+\lambda(1+x)\left(3+\frac{2\lambda}{(1+2\lambda x)(1+\lambda x)}\right)\right]^{-1/2}$$

(94)

$$\Omega^i(x) = \Omega_q$$

$$= \Omega_i \sqrt{\frac{x}{1+x} + \lambda x\left(3+\frac{2\lambda}{(1+2\lambda x)(1+\lambda x)}\right)}$$

(95)

$$B^i(x) = 1 - F_0(\mathbf{q})$$

$$= \left(\frac{(\Omega^i(x)/\Omega_i)^2}{(\Omega^i(x)/\Omega_i)^2 - \lambda x}\right)\left(\frac{1}{\lambda x + \lambda + 1}\right)$$

(96)

If $\lambda \ll 1$, eg, high-$Z$ ions and/or degenerate electrons, then

$$\Omega^i(x) \simeq \Omega_i \sqrt{\frac{x}{1+x}}$$

(97)

However, in the limit of weak coupling ($e^2 \to 0$ or $Z \to 0$) equations (57) and (96), unlike (97), describe ordinary longitudinal acoustic modes with sound velocity $\sqrt{\gamma_i T/m_i} = \sqrt{3T/m_i}$ in which, curiously, $\gamma_i = 3$ corresponds to the longitudinal sound speed in a *solid* with zero Poisson's ratio [17], rather than in a monatomic gas, for which $\gamma_i = 5/3$.

Equation (93) yields the Coulomb logarithm as

$$\ln \Lambda = \lambda \frac{\Omega_i}{2T}(1+e^{-\eta})\int_0^\infty \tilde{B}^i(x)\left(\frac{x}{1+x}\right)^{3/2} \frac{\exp(\Omega^i(x)/T)}{(\exp(\Omega^i(x)/T)-1)^2} \ln\left(\frac{1+e^{\eta-\Delta(x)/T}}{1+e^{\eta-\Delta(x)/T-\Omega^i(x)/T}}\right) dx \quad (98)$$

In the strong-coupling regime, it is generally appropriate to impose a finite upper limit on the wavenumbers contributing to the integrals (93) - (98), similar to that in the Debye model of a crystalline solid, as discussed in section 6.

Both the result (98) and the generalized Ziman formula, as expressed variously by (83), (43) or (44), follow from the general formula, (38), which was derived from the Lenard-Balescu collision integral. However, in the case of the former, the ion structure factor is taken to be given (approximately) by (47) in conjunction with (53); while the electron structure factor at $\omega = \Omega_q$ is taken to be that given by the RPA at $\omega = 0$. The Ziman formula and its generalizations make no detailed assumptions about the ion-structure factor, but rather instead approximate the electron structure factor by its (RPA) form in the vicinity of $\omega = 0$. This results in somewhat simpler formulae, both in terms of representation and evaluation. Also, the lack of assumptions about the ion structure factor provides increased generality in respect of being able to treat strong correlations where the appropriate static structure factor is known. The Ziman formula and its generalizations that depend only upon the static ion structure however do not extend to low temperatures, where either the formula (65), which depends on the dynamic ion structure factor, or (98), must be used instead. The low temperature dense plasma regime is specifically considered in section 6.

Conditions for the validity of the Ziman formulae are examined in more detail in the next section.

## 8 VALIDITY OF THE ZIMAN CONDUCTIVITY FORMULA

### 8.1 The Ziman approximation re-examined

The derivation of the low-temperature form of the Ziman formula, (6), in section 2 depended solely on the validity of the static approximation, but yielded the result in terms of the collective ground-state structure factor (which is the structure factor at $T \ll \Omega_i$). However the approximations made in respect of the electron structure factor, embodied in (5), have been shown generally not to hold in this regime.

We now examine the approximation, as represented by (40), that is at the heart of the generalized Ziman formula.

The heuristic argument already presented is that the ion structure factor should be sufficiently concentrated at low frequencies, for all contributing values of **q**, such that its integral over frequency is virtually exhausted in the region over which the electron structure factor is adequately represented by its $\omega \to 0$ limit. Accordingly, let us start by examining the free-particle structure-factor profiles, $S^0(\mathbf{q}, \omega)$ in $\omega$-space. In the case of a Maxwellian distribution, $S^0(\mathbf{q}, \omega)$ is given by (25), for example. This is a Gaussian distribution centred at $\bar{\omega} = q^2/2m$ and with a standard deviation $= q\sqrt{T/m} = qv$, where $m$ is the particle mass and $v = \sqrt{T/m}$ is the thermal velocity. The overall extent of the function is therefore $q^2/2m + \xi qv$ where $\xi \gtrsim 1$. In the case of extreme degeneracy ($T/T_F \to 0$) the structure factor extends to $q^2/2m + qv_F$ and is zero thereafter [25]. In general, the free-particle structure factor can be taken to extend as far as

$$\hat{\omega} = \frac{q^2}{2m} + q\sqrt{\frac{3T_B}{2m}} \quad (99)$$

which satisfies both criteria. For the approximation being considered to be valid, it is apparent that the following conditions need be satisfied (it being assumed throughout that $m_i/m_e \gg 1$ and $\Omega_e \gg \Omega_i$):

$$\hat{\omega}_e \gg \hat{\omega}_i$$
$$\hat{\omega}_e \gg \Omega_i \quad (100)$$

The first condition is essential. The second is a sufficient condition for the resonances in the interacting ion structure factor to be confined to the low frequency part of the electron structure factor. Conditions relating to the presence of resonances in the electron structure factor, and their possible contribution to the integral, will be considered later. Now, the integration over **q** is confined, by the behaviour of the structure factors, to $q^2 \lesssim \hat{q}^2 = 12 m_e T_B$ (eg, as per equations (32), (33), (42)). Therefore, by application of (99), the conditions (100) become

$$\frac{T_B}{m_e} \gg \frac{T_i}{m_i}$$

$$T_B \gg \Omega_i \left(2-\sqrt{2}\right)/6 \approx 0.1\,\Omega_i \quad (101)$$

These conditions are satisfied over a wide range of plasma densities and temperatures. Particular regimes where they are *not* satisfied however include cold non-degenerate plasmas, or ionized gases, for which $T_e \lesssim 0.1\Omega_i$. Note that the first of conditions (101) is necessary, while the second is weakly sufficient. Since only Coulomb interactions are considered, an overall requirement is that there should be no neutral atoms.

The necessary and sufficient conditions are expressed by $S_{ee}(\mathbf{q},0) \simeq S_{ee}(\mathbf{q},\Omega_\mathbf{q}^i)$ and $S_{ee}(\mathbf{q},0) \simeq S_{ee}(\mathbf{q},\hat{\omega}_i)$ for $q < \hat{q}$, both of which depend on $m_e \ll m_i$.

Let

$$I_\mathbf{q} = \int_{-\infty}^{\infty} S_{ee}(\mathbf{q},-\omega)\tilde{S}_{ii}(\mathbf{q},\omega)\,d\omega \quad (102)$$

The approximation $I_\mathbf{q} \simeq S_{ee}(\mathbf{q},0)\tilde{S}_{ii}(\mathbf{q})$ can be formally derived for a smooth (analytic) function $S_e(\mathbf{q},\omega)$ by Taylor expansion of $S_e(\mathbf{q},\omega)$ about $\omega=0$. This yields

$$I_q = \int_{-\infty}^{\infty} S_{ee}(\mathbf{q},-\omega)\tilde{S}_{ii}(\mathbf{q},\omega)\,d\omega \simeq \int_{-\infty}^{\infty} \left(S_{ee}(\mathbf{q},0) - \omega S'_{ee}(\mathbf{q},0) + \tfrac{1}{2}\omega^2 S''_{ee}(\mathbf{q},0) + \ldots\right)\tilde{S}_{ii}(\mathbf{q},\omega)\,d\omega \quad (103)$$

By using the KMS relation (64), the first derivative of the electron structure factor at zero frequency is found to be $S'_{ee}(\mathbf{q},0) = S_{ee}(\mathbf{q},0)/2T_e$. Hence, with the aid of the f-sum rule, for $q \leq \hat{q}$,

$$\begin{aligned}
I_q &= \int_{-\infty}^{\infty} S_{ee}(\mathbf{q},-\omega)\tilde{S}_{ii}(\mathbf{q},\omega)\,d\omega \simeq S_{ee}(\mathbf{q},0)\left(\tilde{S}_{ii}(\mathbf{q}) - \frac{q^2}{4m_i T_e} + \ldots\right) \\
&\gtrsim S_{ee}(\mathbf{q},0)\left(\tilde{S}_{ii}(\mathbf{q}) - \frac{\hat{q}^2}{4m_i T_e} + \ldots\right) \\
&= S_{ee}(\mathbf{q},0)\left(\tilde{S}_{ii}(\mathbf{q}) - 3\frac{m_e}{m_i}\frac{T_B}{T_e} + \ldots\right) \\
&\simeq S_{ee}(\mathbf{q},0)\tilde{S}_{ii}(\mathbf{q})
\end{aligned} \quad (104)$$

which holds provided that the plasma is not too degenerate, ie

$$\frac{T_B}{T_e} \ll \frac{m_i}{m_e} \quad (105)$$

The condition (105) is not embodied in (101) and is a result of the constraints imposed on the structure factors by the KMS relation and the f-sum rule. Equation (104) does however provide a correction for cases when (105) might not be well satisfied. The condition,

$$T \gg \Theta \quad (106)$$

derived in section 6, for the strongly-coupled degenerate regime, where $\Theta \simeq KD_e\Omega_i$ is the effective Debye temperature for the ion-acoustic modes, and $K$ is the Debye radius, can be expressed as

$$\frac{T_i}{T_F} \gg 2\left(\frac{m_e}{m_i}\right)^{1/2}\left(\frac{4Z}{27}\right)^{1/6} \quad (107)$$

and is generally implied by (105) for $T_e \approx T_i$.

Another approach to the relation (40) is provided by representing the structure factors as Gaussian functions of the frequency in the manner of

$$S(\mathbf{q},\omega) = S(\mathbf{q})\sqrt{\frac{1}{2\pi a}}\exp\left(-\frac{(\omega-b)^2}{2a}\right) \quad (108)$$

in which $a/2b = T$. Such is the case for weakly-interacting Boltzmann distributions (See equation (25)), as well as degenerate plasmas at high frequencies, $\omega$ [26], but may also be true for correlated systems with a Boltzmann-like velocity distribution. Gaussian functions may be used generally to represent the non-collective part of the structure factor.

It is then straightforward to show that

$$I_\mathbf{q} = \int_{-\infty}^{\infty} S_{ee}(\mathbf{q},-\omega)\tilde{S}_{ii}(\mathbf{q},\omega)\,d\omega = S_{ee}(\mathbf{q})\tilde{S}_{ii}(\mathbf{q})\sqrt{\frac{1}{2\pi(a_i+a_e)}}\exp\left(-\frac{(b_i+b_e)^2}{2(a_i+a_e)}\right) \quad (109)$$

so, if $a_e \gg a_i$ and $b_e \gg b_i$,

$$I_\mathbf{q} = \int_{-\infty}^{\infty} S_{ee}(\mathbf{q},-\omega)\tilde{S}_{ii}(\mathbf{q},\omega)\,d\omega \simeq S_{ee}(\mathbf{q})\tilde{S}_{ii}(\mathbf{q})\sqrt{\frac{1}{2\pi a_e}}\exp\left(-\frac{b_e^2}{2a_e}\right)$$

$$= S_{ee}(\mathbf{q},0)\tilde{S}_{ii}(\mathbf{q}) \quad (110)$$

For a plasma characterised by Boltzmann-like distributions of both ions and electrons, as per equations (25) and (212), $a_e/a_i = \mathcal{O}(m_i T_e / m_e T_i)$ and $b_e/b_i = \mathcal{O}(m_i/m_e)$, which satisfy the required conditions, consistently with the first of (101), for such a system. This proof can be generalized to situations when the structure factors are representable by sums of Gaussian functions of the form of (108), provided that the inequalities hold for every pairwise choice of terms, and can thereby be extended to regimes of not-too-extreme degeneracy.

### 8.2  Electron plasmon contribution to the electron-ion collision frequency

The above derivations show that the approximation (40) - (41) should hold for any reasonably well-behaved function $S_{ee}(\mathbf{q},\omega)$. However the analysis has so far ignored possible contributions from the plasmon poles of $1/\varepsilon_e(\mathbf{q},\omega)$. These poles represent the collective resonances in the electron sub-system and are outside the scope of the above analysis. Any Taylor series expansion of $S_{ee}(\mathbf{q},\omega)$ about $\omega=0$ would not extend beyond the neighbourhood of such a pole, which requires that any poles that restrict the circle of convergence in the vicinity of the real $\omega$-axis to being within the range of $\tilde{S}_{ii}(\mathbf{q},\omega)$ should be subtracted out and dealt with separately. The contribution to $I_\mathbf{q}$ from these poles will now be considered.

To illustrate this, we use the generalized plasmon-pole approximation [17]

$$S_{ee}(\mathbf{q},\omega) = A_\mathbf{q}^e \delta_\mathbf{q}^e(\omega) + B_\mathbf{q}^e \frac{q^2}{2m_e} g_\mathbf{q}^e(\omega) \quad (111)$$

The first term is the regular independent-particle term, which is treated as above. The second term is the electron plasmon term, which contains the resonance poles and which, so far, has not been taken into account. In the weak damping limit, this is represented by

$$g_{\mathbf{q}}^e(\omega) = \frac{1}{\Omega_{\mathbf{q}}^e}\left[(\mathcal{N}_{\mathbf{q}}^e + 1)\delta(\omega - \Omega_{\mathbf{q}}^e) + \mathcal{N}_{\mathbf{q}}^e \delta(\omega + \Omega_{\mathbf{q}}^e)\right] \tag{112}$$

where $\mathcal{N}_{\mathbf{q}}^e$ is the Bose-Einstein function,

$$\mathcal{N}_{\mathbf{q}}^e = \frac{1}{\exp(\Omega_{\mathbf{q}}^e/T_e) - 1} \tag{113}$$

The electron-plasmon contribution to (102) is therefore

$$\begin{aligned}
I_{\mathbf{q}}^{\text{plasmon}} &= B_{\mathbf{q}}^e \frac{q^2}{2m_e} \int_{-\infty}^{\infty} g_{\mathbf{q}}^e(-\omega)\, \tilde{S}_{ii}(\mathbf{q},\omega)\, d\omega \\
&= B_{\mathbf{q}}^e \frac{q^2}{2m_e \Omega_{\mathbf{q}}^e} \int_{-\infty}^{\infty}\left[(\mathcal{N}_{\mathbf{q}}^e+1)\delta(\omega+\Omega_{\mathbf{q}}^e) + \mathcal{N}_{\mathbf{q}}^e \delta(\omega - \Omega_{\mathbf{q}}^e)\right]\tilde{S}_{ii}(\mathbf{q},\omega)\, d\omega \\
&= B_{\mathbf{q}}^e \frac{q^2}{2m_e \Omega_{\mathbf{q}}^e}\left[(\mathcal{N}_{\mathbf{q}}^e+1)\tilde{S}_{ii}(\mathbf{q},-\Omega_{\mathbf{q}}^e) + \mathcal{N}_{\mathbf{q}}^e \tilde{S}_{ii}(\mathbf{q},\Omega_{\mathbf{q}}^e)\right] \\
&= B_{\mathbf{q}}^e \frac{q^2}{2m_e \Omega_{\mathbf{q}}^e}\left[(\mathcal{N}_{\mathbf{q}}^e+1)\exp(-\Omega_{\mathbf{q}}^e/T_i) + \mathcal{N}_{\mathbf{q}}^e\right]\tilde{S}_{ii}(\mathbf{q},\Omega_{\mathbf{q}}^e)
\end{aligned} \tag{114}$$

where use has been made of the KMS relation, $\tilde{S}_{ii}(\mathbf{q},-\omega) = \exp(-\omega/T_i)\tilde{S}_{ii}(\mathbf{q},\omega)$. Substituting for $\mathcal{N}_{\mathbf{q}}^e$ according to (113) yields

$$I_{\mathbf{q}}^{\text{plasmon}} = B_{\mathbf{q}}^e \frac{q^2}{2m_e \Omega_{\mathbf{q}}^e}\left(\frac{\exp\left(\Omega_{\mathbf{q}}^e\left(\frac{1}{T_e} - \frac{1}{T_i}\right)\right) + 1}{\exp\left(\frac{\Omega_{\mathbf{q}}^e}{T_e}\right) - 1}\right) \tilde{S}_{ii}(\mathbf{q},\Omega_{\mathbf{q}}^e) \tag{115}$$

Now, from (36), using the high-frequency asymptotic form of $\tilde{\varepsilon}_i$, since $\Omega_{\mathbf{q}}^e \gg \Omega_i$

$$\tilde{S}_{ii}(\mathbf{q},\Omega_{\mathbf{q}}^e) = \frac{S_{ii}^0(\mathbf{q},\Omega_{\mathbf{q}}^e)}{|\tilde{\varepsilon}_i(\mathbf{q},\Omega_{\mathbf{q}}^e)|^2} = \frac{S_{ii}^0(\mathbf{q},\Omega_{\mathbf{q}}^e)}{\left(1 - (\Omega_i/\Omega_{\mathbf{q}}^e)^2\right)^2} \tag{116}$$

where $S_{ii}^0(\mathbf{q},\omega)$ is the dynamic structure factor for a non-interacting Boltzmann gas, which is given by (25), whereupon, ignoring terms of $\mathcal{O}(\Omega_i^2/\Omega_e^2) = \mathcal{O}(m_e/m_i)$,

$$\tilde{S}_{ii}(\mathbf{q},\Omega_{\mathbf{q}}^{e}) \simeq \sqrt{\frac{m_i}{2\pi q^2 T_i}} \exp\left(-\frac{m_i}{2q^2 T_i}\left(\Omega_{\mathbf{q}}^{e} - \frac{q^2}{2m_i}\right)^2\right)$$

$$= \sqrt{\frac{m_i}{2\pi q^2 T_i}} \exp\left(-\frac{\Delta\Omega_i^{e}}{T_i}\right) \tag{117}$$

where

$$\Delta\Omega_i^{e} = \frac{m_i}{2q^2}\left(\Omega_{\mathbf{q}}^{e} - \frac{q^2}{2m_i}\right)^2 = \left(\frac{\Omega_{\mathbf{q}}^{e}}{4}\right)\left(\frac{2m_i\Omega_{\mathbf{q}}^{e}}{q^2}\right)\left(1 - \frac{q^2}{2m_i\Omega_{\mathbf{q}}^{e}}\right)^2 \tag{118}$$

In terms of $I_{\mathbf{q}}$ defined by (102), the collision frequency (38) is given by

$$\bar{\nu} = \frac{\pi n_e n_i}{4 m_e \langle k^3 \rangle} \frac{1+e^{-\eta}}{T_e} \int q^2 |V(\mathbf{q})|^2 I_{\mathbf{q}} \frac{d^3 \mathbf{q}}{4\pi}$$

$$= \nu_0 \frac{\pi^2 n_e}{m_e^2 T_e}(1+e^{-\eta}) \int_0^\infty I_{\mathbf{q}}\, dq \tag{119}$$

where $\nu_0$ is the standard collision frequency given by (81).

The contribution to the Coulomb logarithm, $\ln\Lambda = \bar{\nu}/\nu_0$, from electron plasma modes is therefore

$$\ln\Lambda^{\text{plasmon}} = \frac{\pi^2 n_e}{m_e^2 T_e}(1+e^{-\eta})\int_0^\infty I_{\mathbf{q}}^{\text{plasmon}}\, dq$$

$$= \left(\frac{\pi}{2}\right)^{3/2}\left(\frac{m_i}{T_i}\right)^{1/2}\frac{n_e \Omega_e}{m_e^2 T_e T_B}(1+e^{-\eta})\int_0^\infty F(x)\, dx \tag{120}$$

where the integration variable has been changed to $x = q^2 D_e^2$, and

$$F(x) = \frac{1}{2}\frac{B_e(x)}{\sqrt{G_e(x)}}\left(\frac{\exp\left(\Omega_{\mathbf{q}}^{e}\left(\frac{1}{T_e}-\frac{1}{T_i}\right)\right)+1}{\exp\left(\frac{\Omega_{\mathbf{q}}^{e}}{T_e}\right)-1}\right)\exp\left(-\frac{\Delta\Omega_i^{e}}{T_i}\right) \tag{121}$$

where

$$G_e(x) = \left(\Omega_{\mathbf{q}}^{e}/\Omega_e\right)^2 \simeq 1+3x \tag{122}$$

and, from equations (165) of ref. [17],

$$B_e(x) = B_q^e = \left(\frac{G_e(x)}{G_e(x) - x}\right)\left(\frac{1}{1+x}\right) \qquad (123)$$

Using the ideal electron gas formula (26), equation (120) reduces to

$$\ln \Lambda^{\text{plasmon}} = \frac{1}{2\sqrt{\pi}}\left(\frac{m_i T_e}{m_e T_i}\right)^{1/2} \frac{\Omega_e}{T_B} I_{1/2}(\eta)\left(1+e^{-\eta}\right)\int_0^\infty F(x)\,dx \qquad (124)$$

which, for a non-degenerate (Maxwellian) plasma, becomes

$$\ln \Lambda^{\text{plasmon}} \xrightarrow[\eta \to -\infty]{} \frac{1}{4}\left(\frac{m_i T_e}{m_e T_i}\right)^{1/2} \frac{\Omega_e}{T_e} \int_0^\infty F(x)\,dx \qquad (125)$$

while, in the totally degenerate limit,

$$\ln \Lambda^{\text{plasmon}} \xrightarrow[\eta \to \infty]{} \frac{1}{2\sqrt{\pi}}\left(\frac{m_i T_F}{m_e T_i}\right)^{1/2} \frac{\Omega_e}{T_e} \int_0^\infty F(x)\,dx \qquad (126)$$

The integral $\int_0^\infty F(x)\,dx$ is dominated by the factor $\exp(-\Delta\Omega_i^e/T_i)$ in the integrand, with the dominant contribution to the integral coming from the region of the saddle point at[c]

$$x_0 = 2m_i \Omega_q^e D_e^2 = 2\frac{m_i T_B}{m_e \Omega_e}\sqrt{G_e(x_0)} \qquad (127)$$

With $G_e(x_0)$ given according to (122), this yields

$$x_0 = 6\left(\frac{m_i T_B}{m_e \Omega_e}\right)^2 \left(1 + \sqrt{1 + \left(\frac{m_e \Omega_e}{3m_i T_B}\right)^2}\right) \qquad (128)$$

so, for $m_e \Omega_e \gg m_i T_B$,

$$x_0 \approx \frac{2m_i T_B}{m_e \Omega_e} \ll 1 \qquad (129)$$

or, for $m_e \Omega_e \ll m_i T_B$

$$x_0 \approx 12\left(\frac{m_i T_B}{m_e \Omega_e}\right)^2 \gg 1 \qquad (130)$$

In particular, the regime of weak damping corresponds to

$$m_i T_B \leq \tfrac{1}{4} m_e \Omega_e \quad \Rightarrow \quad x_0 \leq 1 \tag{131}$$

---

[c] If $\partial \ln G_e(x)/\partial \ln x < 1 \quad \forall x > 0$, then this is the only saddle point that exists in $x > 0$.

In terms of these quantities, and provided that, $m_i \Omega_e^2 D_e^2 / T_i = (m_i T_B)/(m_e T_i) \gg 1$, which is generally true, the integral can be evaluated approximately as follows

$$\int_0^\infty F(x) \, dx \sim \frac{x_0}{2\gamma_0} B_e(x_0) \left( \frac{\exp\left[\gamma_0 \Omega_e \left(\frac{1}{T_e} - \frac{1}{T_i}\right)\right] + 1}{\exp\left(\gamma_0 \frac{\Omega_e}{T_e}\right) - 1} \right) \int_0^\infty \exp\left(-\gamma_0 \frac{\Omega_e}{4T_i} \frac{(1-t)^2}{t}\right) dt$$

$$= \frac{x_0}{2\gamma_0} B_e(x_0) \left( \frac{\exp\left[\gamma_0 \Omega_e \left(\frac{1}{T_e} - \frac{1}{T_i}\right)\right] + 1}{\exp\left(\frac{\gamma_0 \Omega_e}{T_e}\right) - 1} \right) \exp\left(\gamma_0 \frac{\Omega_e}{2T_i}\right) K_1\left(\gamma_0 \frac{\Omega_e}{2T_i}\right) \tag{132}$$

where $\gamma_0 \equiv \sqrt{G_e(x_0)} = x_0 m_e \Omega_e / 2 m_i T_B$ and where $K_1(x)$ denotes a modified Bessel function of the second kind [27]. Hence, combining this result with (124),

$$\ln \Lambda^{\text{plasmon}} \simeq \frac{1}{\sqrt{\pi}} \left(\frac{m_i}{m_e}\right)^{3/2} \left(\frac{T_e}{T_i}\right)^{1/2} I_{1/2}(\eta)(1 + e^{-\eta}) B_e(x_0) \exp\left(\gamma_0 \frac{\Omega_e}{2T_i}\right) K_1\left(\gamma_0 \frac{\Omega_e}{2T_i}\right)$$

$$\times \left( \frac{\exp\left[\gamma_0 \Omega_e \left(\frac{1}{T_e} - \frac{1}{T_i}\right)\right] + 1}{\exp\left(\frac{\gamma_0 \Omega_e}{T_e}\right) - 1} \right) \tag{133}$$

At low ion temperatures, such that $\gamma_0 \Omega_e \gg T_i$, this becomes

$$\ln \Lambda^{\text{plasmon}} \sim \left(\frac{m_i}{m_e}\right)^{3/2} \left(\frac{T_e}{\gamma_0 \Omega_e}\right)^{1/2} I_{1/2}(\eta)(1 + e^{-\eta}) B_e(x_0) \left( \frac{\exp\left[\gamma_0 \Omega_e \left(\frac{1}{T_e} - \frac{1}{T_i}\right)\right] + 1}{\exp\left(\frac{\gamma_0 \Omega_e}{T_e}\right) - 1} \right) \tag{134}$$

which is negligible if $T_i \leq T_e \ll \tfrac{2}{3} \gamma_0 \Omega_e / \ln(m_i/m_e)$. However, at such temperatures, the excitation of the electron plasmon modes is generally negligible anyway.

A necessary condition for (133) *not* to be negligible is therefore $T_e \gtrsim \gamma_0 \Omega_e / \ln(m_i/m_e)$, which, since $\gamma_0 \geq 1$ and $T_B \geq T_e$, implies $m_i T_B \gg m_e \Omega_e$ and hence, according to (130), $x_0 \gg 1$, whereupon, from (123), $B_e(x_0) \approx 3/2 x_0$ and hence

$$\ln \Lambda^{\text{plasmon}} \simeq \frac{1}{8\gamma_0 \sqrt{\pi}} \left(\frac{m_e}{m_i}\right)^{1/2} \left(\frac{T_e T_i}{T_B^2}\right)^{1/2} \left(\frac{\Omega_e}{T_B}\right) I_{1/2}(\eta)(1+e^{-\eta}) \left(\gamma_0 \frac{\Omega_e}{2T_i}\right) \exp\left(\gamma_0 \frac{\Omega_e}{2T_i}\right) K_1\left(\gamma_0 \frac{\Omega_e}{2T_i}\right)$$
$$\times \left(\frac{\exp\left(\gamma_0 \Omega_e \left(\frac{1}{T_e} - \frac{1}{T_i}\right)\right) + 1}{\exp\left(\frac{\gamma_0 \Omega_e}{T_e}\right) - 1}\right) \quad (135)$$

If moreover the ion temperature is high, such that $T_i \gg \gamma_0 \Omega_e$, then this becomes

$$\ln \Lambda^{\text{plasmon}} \simeq \frac{1}{8\gamma_0 \sqrt{\pi}} \left(\frac{m_e}{m_i}\right)^{1/2} \left(\frac{T_e T_i}{T_B^2}\right)^{1/2} \left(\frac{\Omega_e}{T_B}\right) I_{1/2}(\eta)(1+e^{-\eta}) \left(\frac{\exp\left(\gamma_0 \Omega_e \left(\frac{1}{T_e} - \frac{1}{T_i}\right)\right) + 1}{\exp\left(\frac{\gamma_0 \Omega_e}{T_e}\right) - 1}\right) \quad (136)$$

which remains negligible at all reasonable ion temperatures. However this has assumed that (123) can be extrapolated to large $x$. Nevertheless it is generally reasonable to assume that, $B_e(x) \leq 3/2x$, and $\gamma_0 \geq 1$, whereupon, for $T_i \gg \gamma_0 \Omega_e$, using that $(T_B/T_e)^{3/2} I_{1/2}(\eta)(1+e^{-\eta}) \sim 1$,

$$\ln \Lambda^{\text{plasmon}} \leq \frac{1}{8\gamma_0 \sqrt{\pi}} \left(\frac{m_e}{m_i}\right)^{1/2} \left(\frac{T_e T_i}{T_B^2}\right)^{1/2} \left(\frac{\Omega_e}{T_B}\right) I_{1/2}(\eta)(1+e^{-\eta}) \left(\frac{\exp\left(\gamma_0 \Omega_e \left(\frac{1}{T_e} - \frac{1}{T_i}\right)\right) + 1}{\exp\left(\frac{\gamma_0 \Omega_e}{T_e}\right) - 1}\right)$$

$$\sim \frac{1}{8\gamma_0 \sqrt{\pi}} \left(\frac{m_e T_i}{m_i T_e}\right)^{1/2} \left(\frac{\Omega_e}{T_e}\right) \leq \frac{1}{8\gamma_0^2 \sqrt{\pi}} \left(\frac{m_e T_i}{m_i T_e}\right)^{1/2} \ln\left(\frac{m_i}{m_e}\right) \quad (137)$$

$$< \frac{1}{8\sqrt{\pi}} \left(\frac{m_e T_i}{m_i T_e}\right)^{1/2} \ln\left(\frac{m_i}{m_e}\right)$$

For this not to be negligible would require $T_i \gtrsim (m_i/m_e) T_e$, which would imply, for Maxwellian distributions, that the mean ion thermal velocity is greater than that of the electrons.

If, on the other hand, $T_i \ll \gamma_0 \Omega_e$, then (135) becomes

$$\ln \Lambda^{\text{plasmon}} \simeq \frac{1}{16} \left(\frac{m_e}{m_i}\right)^{1/2} \left(\frac{T_e}{\gamma_0 T_B}\right)^{1/2} \left(\frac{\Omega_e}{T_B}\right)^{3/2} I_{1/2}(\eta)(1+e^{-\eta}) \left(\frac{\exp\left(\gamma_0 \Omega_e \left(\frac{1}{T_e} - \frac{1}{T_i}\right)\right) + 1}{\exp\left(\frac{\gamma_0 \Omega_e}{T_e}\right) - 1}\right) \quad (138)$$

while the more general inequality (137) becomes replaced by

$$\ln \Lambda^{\text{plasmon}} \leq \frac{1}{16}\left(\frac{m_e}{m_i}\right)^{1/2}\left(\frac{T_e}{\gamma_0 T_B}\right)^{1/2}\left(\frac{\Omega_e}{T_B}\right)^{3/2} I_{1/2}(\eta)\left(1+e^{-\eta}\right)\left(\frac{\exp\left(\gamma_0 \Omega_e\left(\frac{1}{T_e}-\frac{1}{T_i}\right)\right)+1}{\exp\left(\frac{\gamma_0 \Omega_e}{T_e}\right)-1}\right)$$

(139)

$$< \frac{1}{16}\left(\frac{m_e}{m_i}\right)^{1/2}\left(\frac{\left(\frac{\Omega_e}{T_e}\right)^{3/2}}{\exp\left(\frac{\gamma_0 \Omega_e}{T_e}\right)-1}\right)$$

both of which are likely to be small given that $m_e \ll m_i$.

To summarise, the generalized Ziman formula has been found to be generally restricted, in its validity, to temperatures satisfying

$$\frac{m_i}{m_e} T_e \gg T_B \gg \frac{m_e}{m_i} T_i$$

$$T_B \gg 0.1 \Omega_i$$

(140)

while the standard formula, requires, in addition, that the plasma be degenerate, ie $T_F \gg T_e$.

It is concluded that the approximation represented by (40) has wide validity, extending from the hot liquid metal regime to non-relativistic hot plasmas in local thermodynamic equilibrium (LTE) at virtually any temperature and density. All the regimes where (140) are not true are associated with an electron dynamic structure factor that is narrow on the scale of the ion dynamic structure factor.

## 9  CONSIDERATION OF ELECTRON-ELECTRON COLLISIONS

### 9.1  Non-Lorentzian corrections

It is perhaps a common misconception that the Lorentzian model amounts to treating electron-electron (ee) collisions as if they do not occur. In fact, the Lorentzian model implicitly assumes that the electrons remain close to LTE with the current flow occurring isothermally. For this to be so, the electrons must interact strongly with each other and/or with their surroundings. Also, while it is clear that electron-electron collisions alone cannot give rise to any resistivity, due to momentum being conserved in the electron subsystem, this is insufficient reason to neglect them.

Electron-electron collisions affect the conductivity by redistributing the electron energies. There is an extensive literature [11], [12], [28], [7], [29] dealing with this problem in the context of non-LTE classical transport theory based upon the Chapman-Enskog (CE) approach [30]. This method uses a first-principles approach to solving the Boltzmann equation involving the non-linear electron-electron scattering terms. Apart from being unwieldy, this method suffers from a number of problems: the fact that it does not align with the Lorentzian approximation makes it difficult to reconcile the two approaches. In particular, the Lorentzian (Ziman) model incorporates ion-ion correlations as well as a more consistent treatment of the plasma screening. These may be down to the fact that the Lorentzian approach does permit and thereby encompass electron-electron scattering at some level of approximation in various parameter domains, while the treatment of the electrons in the CE model is fundamentally classical. Other methods include density functional theory (DFT) [31] and molecular dynamics (MD) combined with standard linear response theory [32], [33]. A contemporary review of the ee scattering problem in the context of the generalized linear response theory (GLRT) of Zubarev *et al* [34],[35] is given in [36].

A solution to the problem of treating electron-electron collisions was first given by Spitzer *et al* [37], [28], [11], [12]. Their result is expressed by the formula

$$\bar{v} = \frac{\bar{v}_{\rm L}}{\gamma_{\rm e}(Z)} \tag{141}$$

where $\bar{v}$ is the effective collision frequency defined by (7) and $\bar{v}_{\rm L}$ is the same parameter given in the Lorentzian approximation, with $\gamma_{\rm e}(Z)$ given by:

$$\gamma_{\rm e}(Z) = \frac{1}{\exp\left(1-(1-\ln 2)\left((Z-1)/Z\right)^{a(Z)}\right)-1} \quad , \quad Z > 1$$

$$= \frac{1}{e-1} \quad , \quad Z \le 1 \tag{142}$$

in which $a(Z) = 1.07545 + .68425\sqrt{\ln Z}$, which is a parameterization based on the numerical values given in [11]. A Puiseux series expansion of (142) in powers of $1/Z$ yields

$$\gamma_{\rm e}(Z) = 1 + 2(\ln 2 - 1)\frac{a(Z)}{Z} + \mathcal{O}\left(\frac{1}{Z^2}\right) \tag{143}$$

An approach, which allows the electrons to be treated quantum mechanically, and which incorporates the non-linear electron-electron scattering terms in a consistent way, is described in APPENDIX A.2. This gives the following formula for the effective collision frequency $\bar{v}$,

$$\bar{v} = \frac{\sum_\beta \dfrac{v_\beta^{\text{ei}}}{v_\beta^{\text{ei}} + v_\beta^{\text{ee}}} \varepsilon_\beta p_\beta^0 q_\beta^0}{\sum_\beta \dfrac{1}{v_\beta^{\text{ei}} + v_\beta^{\text{ee}}} \varepsilon_\beta p_\beta^0 q_\beta^0} \tag{144}$$

where $v_\beta^{\text{ei}}$ is the electron-ion collision frequency as given by the Lorentzian model and $v_\beta^{\text{ee}}$ is the electron-electron collision frequency, which is assumed to be derivable independently of the electron-ion collisions. Note that, if $v_\beta^{\text{ei}} = 0$, $\forall \beta$, then $\bar{v} = 0$, which implies that electron-electron collisions alone do not give rise to any resistivity. Making the Z-dependence explicit by writing

$$\frac{v_\beta^{\text{ee}}}{v_\beta^{\text{ei}}} = \frac{b_\beta}{Z} \tag{145}$$

then, if $b_\beta = 0$, $\forall \beta$ or if $b_\beta = b$ is a finite constant independent of $\beta$, equation (144) reduces to $\bar{v} = \bar{v}_L \equiv \sum_\beta \varepsilon_\beta p_\beta^0 q_\beta^0 / \sum_\beta (1/v_\beta^{\text{ei}}) \varepsilon_\beta p_\beta^0 q_\beta^0$. In general, $b_\beta$ in (145) may retain a slow logarithmic dependence on $Z$, as is the case in (143).

Detailed calculations [38] of weakly-coupled non-degenerate plasmas using GLRT, neglecting ion-ion correlations, show that $b_\beta$ increases monotonically with wavenumber, $k_\beta$, for $k_\beta D < 50$ levelling off at $b_\beta \simeq 0.45$ for $k_\beta D \gtrsim 50$. Degeneracy alone would be expected further to reduce $v_\beta^{\text{ee}}$ relative to $v_\beta^{\text{ei}}$ because of Pauli blocking of the states of the scattering electrons in addition to those of the scattered electrons. Otherwise the ratio of the collision frequencies is generally not constant, chiefly because of correlations, which are most important at low temperatures, and other differentiating effects incorporated in the Coulomb Logarithm.

Making the assumption that $b_\beta/Z < 1$, leads to corrections to the Lorentzian model in the form of a power (Puiseux) series in $1/Z$ as follows

$$\frac{\bar{v}_L}{\bar{v}} = \frac{\bar{\tau}}{\bar{\tau}_L} = 1 - \frac{A_1}{Z} + \frac{A_2}{Z^2} - \frac{A_3}{Z^3} + \ldots \tag{146}$$

where the coefficients $A_r$ depend upon the differences between differently weighted averages of powers of $b_\beta$ as given by

$$\overline{b^n} = \frac{\sum_\beta \varepsilon_\beta p_\beta q_\beta b_\beta{}^n}{\sum_\beta \varepsilon_\beta p_\beta q_\beta} = \frac{2}{3T} \sum_\beta \varepsilon_\beta p_\beta q_\beta b_\beta{}^n$$

(147)

$$\overline{\overline{b^n}} = \frac{\sum_\beta \varepsilon_\beta{}^{5/2} p_\beta q_\beta b_\beta{}^n}{\sum_\beta \varepsilon_\beta{}^{5/2} p_\beta q_\beta}$$

according to the formulae

$$A_1 = \overline{\overline{b}} - \overline{b}$$

(148)

$$A_n = \overline{\overline{b^n}} - \overline{b^n} - \sum_{r=1}^{n-1} A_r \overline{b^{n-r}}$$

in which the coefficients $A_n$ are defined by the general series expansion,

$$\frac{1 + \sum_{n=1}^{\infty} \overline{\overline{b^n}} x^n}{1 + \sum_{n=1}^{\infty} \overline{b^n} x^n} = 1 + \sum_{n=1}^{\infty} A_n x^n$$

(149)

Given that $b_\beta$ is an increasing function of the energy ($\partial b_\beta / \partial \varepsilon_\beta \geq 0$) the energy weightings in (147) imply that $\overline{\overline{b^n}} > \overline{b^n}$ and hence that $A_1 > 0$. For sufficiently large $Z$, this yields $\overline{V}_L < \overline{V}$ and hence $\sigma_L > \sigma$ [29]. Given that the coefficients $b_\beta$ retain a possible logarithmic dependence on $Z$ as therefore do the coefficients $A_n$, equation (146) is, in general, a Puiseux series.

For a completely degenerate system,

$$\overline{\overline{b^n}} = \overline{b^n} = b_F{}^n = b^n(\varepsilon_F)$$

(150)

where $b_F$ is the ratio of the Coulomb logarithms evaluated at the Fermi energy, which yields $A_n = 0$, $\forall n$. The model therefore predicts that the Lorentzian approximation should hold well for highly degenerate systems [36].

## 10 CONCLUSIONS

This report has examined some generalizations of the Ziman formula for the electrical conductivity of a liquid metal. These generalizations extend to hot Lorentzian plasmas as well as to systems of arbitrary electron degeneracy, and also resolve the inconsistencies, between various formulae for the conductivity based upon the Ziman formula, identified in [7]. Consideration of the Lenard-Balescu collision integral, in which finite amounts of energy as well as momentum are exchanged in electron-ion collisions, leads to the main results of this article, which are the results for the collision frequency and electrical conductivity expressed by equations (43) - (46). Using the ion structure factor from ref. [17], leads to new formulae for the electrical conductivity.

Ion plasma modes are implicitly taken into account, and do not need to be treated separately, except where the approximation (40) is not valid, which would be when $S_{ee}(\mathbf{q},0) \neq S_{ee}(\mathbf{q},\Omega_{\mathbf{q}}^i)$ for $q \lesssim \sqrt{12 m_e T_B}$. This situation is addressed in section 6. Electron plasma modes are however neglected in making the approximation (40), owing to the electron structure factor being replaced by its zero frequency limit. However this can be corrected for in the manner described in section 8.2. Indications are that this correction is generally negligible.

The formalism also provides explicit formulae for the ion structure factor in terms of the dielectric functions. These lead to closed form expressions for the Coulomb logarithm as described in section 7.

The basic Ziman formula and its generalizations discussed above, in sections 3-5, do not extend to arbitrarily low temperatures. The generalization of the theory to low-temperature dense highly-degenerate regimes is considered in section 6, the outcome of which is that the result (43) for the effective collision frequency is replaced with that given by (65). In the low temperature regime, for which the results are expressed by (76) and (77), the theory accords with the well-known Gruneisen-Bloch theory of metallic solids, except that, in the case of plasmas, the collective motion is described in terms of the ion-acoustic modes. Therefore, the models presented here represent a unification of various models of the conductivity of plasmas and metals across a wide range of conditions, while resolving some outstanding inconsistencies.

Consideration is given to including an explicit treatment of the effects of electron-electron collisions on the conductivity. Corrections are found in the form of a series expansion in powers of $1/Z$ in which the coefficients are differences between differently weighted powers of the ratio of the ee and ei Coulomb logarithms. These corrections are expected to be small for high-Z plasmas and metals, and to vanish in the high degeneracy limit.

.

# APPENDIX A  THE BOLTZMANN TRANSPORT EQUATION

## A.1    Application to Lorentzian plasmas.

The basic form of the Boltzmann transport equation, henceforth referred to as the Boltzmann equation, for the ensemble-averaged occupancies $\{\, 0 \le p_\beta \le 1 \,\}$ of the electronic states $\{\,\beta\,\}$, is

$$\frac{\partial p_\beta}{\partial t} = \left.\frac{\partial p_\beta}{\partial t}\right|_{\text{field}} + \left.\frac{\partial p_\beta}{\partial t}\right|_{\text{collisions}} = 0 \qquad (151)$$

in which

$$\left.\frac{\partial p_\beta}{\partial t}\right|_{\text{field}} = -\mathbf{f}\cdot\frac{\partial p_\beta}{\partial \mathbf{k}} \qquad (152)$$

gives the response to an external force, $\mathbf{f}$, acting on each electron, and

$$\left.\frac{\partial p_\beta}{\partial t}\right|_{\text{collisions}} = \sum_\alpha \left(\nu_{\beta\alpha} - \nu_{\alpha\beta}\right) \qquad (153)$$

is the net rate of collisional population of the state, being the difference between the electron-ion (ei) collisional rates $\nu_{\alpha\beta}$ into and out of the state $\beta$. Introducing the relaxation times $\tau_\beta$ defined by

$$\left.\frac{\partial p_\beta}{\partial t}\right|_{\text{collisions}} = -\frac{\delta p_\beta}{\tau_\beta} \qquad (154)$$

where $\delta p_\beta = p_\beta - p_\beta^0$ is the deviation from LTE (denoted by $^0$) and taking $p_\beta^0$ to be represented by a Fermi-Dirac distribution, $p^0 = f(\varepsilon;\mu,T) \equiv (1+\exp((\varepsilon-\mu)/T))^{-1}$ equations (151), (152), (154) yield the first-order (linear) response as

$$\delta p_\beta = \tau_\beta \mathbf{f}\cdot\mathbf{v}_\beta \frac{p_\beta^0 q_\beta^0}{T} \qquad (155)$$

where $\mathbf{v} = \partial\varepsilon/\partial\mathbf{k}$ is the electron velocity, $T$ is the electron temperature, and $q^0 = 1 - p^0$. The electric current that flows in response to an external electric field $\mathbf{E}$ is therefore

$$\mathbf{j} = e\frac{1}{\mathfrak{V}}\sum_\beta \mathbf{v}_\beta \delta p_\beta = \frac{e^2}{T}\frac{1}{\mathfrak{V}}\sum_\beta \mathbf{v}_\beta\mathbf{v}_\beta\cdot\mathbf{E}\,\tau_\beta p_\beta^0 q_\beta^0 \qquad (156)$$

which yields the conductivity, defined by $\mathbf{j} = \sigma\mathbf{E}$, as

$$\sigma = \frac{e^2}{T}\frac{1}{\mathfrak{V}}\sum_\beta \mathbf{v}_\beta\mathbf{v}_\beta \tau_\beta p_\beta^0 q_\beta^0 \qquad (157)$$

In an isotropic system, only the diagonal components of $\boldsymbol{\sigma}$ are non-vanishing. The scalar conductivity, in terms of which $\mathbf{j} = \sigma \mathbf{E}$, is therefore

$$\sigma = \frac{e^2}{3T} \frac{1}{\mathfrak{V}} \sum_\beta v_\beta^2 \tau_\beta p_\beta^0 q_\beta^0 \tag{158}$$

It now remains to determine $\tau_\beta$ in terms of the microscopic variables $v_{\alpha\beta}$. Let $v_{\alpha\beta} = w_{\alpha\beta} p_\beta q_\alpha$ where $w_{\alpha\beta}$ is a coefficient that is independent of the state occupancies. Then, expanding around the LTE occupancies,

$$v_{\alpha\beta} = w_{\alpha\beta} \left( p_\beta^0 + \delta p_\beta \right) \left( q_\alpha^0 - \delta p_\alpha \right)$$

$$= v_{\alpha\beta}^0 \left( 1 + \frac{\delta p_\beta}{p_\beta^0} \right) \left( 1 - \frac{\delta p_\alpha}{q_\alpha^0} \right) \tag{159}$$

where $v_{\alpha\beta}^0$ are the LTE rates, which satisfy *detailed balance*

$$v_{\alpha\beta}^0 = v_{\beta\alpha}^0 \tag{160}$$

Combining (159) and (160) yields

$$v_{\beta\alpha} - v_{\alpha\beta} = v_{\alpha\beta}^0 \left( \frac{\delta p_\alpha}{p_\alpha^0 q_\alpha^0} - \frac{\delta p_\beta}{p_\beta^0 q_\beta^0} \right) \tag{161}$$

whereupon, substituting for $\{\delta p_\alpha\}$ from (155),

$$v_{\beta\alpha} - v_{\alpha\beta} = \frac{v_{\alpha\beta}^0}{T} \mathbf{f} \cdot \left( \mathbf{v}_\alpha \tau_\alpha - \mathbf{v}_\beta \tau_\beta \right) \tag{162}$$

Henceforth, we will be working solely in terms of LTE quantities and so can drop the $^0$ superscripts. Upon substituting (162) into (153) and making use of (152), (154) and (155), the Boltzmann equation for a system in or close to LTE, (151), is rendered as

$$\mathbf{f} \cdot \mathbf{v}_\beta p_\beta q_\beta = \mathbf{f} \cdot \sum_\alpha v_{\alpha\beta} \left( \mathbf{v}_\beta \tau_\beta - \mathbf{v}_\alpha \tau_\alpha \right) \tag{163}$$

Since the relaxation times are independent of the arbitrary force $\mathbf{f}$, then, in an isotropic medium, this implies

$$\mathbf{v}_\beta p_\beta q_\beta = \sum_\alpha v_{\alpha\beta} \left( \mathbf{v}_\beta \tau_\beta - \mathbf{v}_\alpha \tau_\alpha \right) \tag{164}$$

or, replacing $\mathbf{v}$ by $\mathbf{k}/m_e$,

$$\mathbf{k}_\beta p_\beta q_\beta = \sum_\alpha v_{\alpha\beta} \left( \mathbf{k}_\beta \tau_\beta - \mathbf{k}_\alpha \tau_\alpha \right) \tag{165}$$

Equation (165) is an algebraic representation of the Boltzmann equation for the transport of fermion particles and provides the fundamental relationship between the LTE collision rates $v_{\alpha\beta} = v_{\beta\alpha}$ and the relaxation times $\tau_\alpha$, $\tau_\beta$, … .

Equation (165) can be solved for an isotropic system in the elastic limit, when the relaxation times depend only upon the (kinetic) energy of the state and not upon the direction of the velocity, and the collision processes involve negligible exchanges of energy. Taking the scalar product with $\mathbf{k}_\beta$ then yields,

$$k_\beta^2 p_\beta q_\beta = \tau_\beta \sum_\alpha v_{\alpha\beta} \left( \mathbf{k}_\beta - \mathbf{k}_\alpha \right) \cdot \mathbf{k}_\beta = \frac{\tau_\beta}{2} \sum_\alpha v_{\alpha\beta} q^2 \tag{166}$$

where $\mathbf{q} = \mathbf{k}_\alpha - \mathbf{k}_\beta$. Hence

$$\frac{1}{\tau_\beta} = v_\beta \equiv \frac{1}{p_\beta q_\beta} \sum_\alpha \frac{q^2}{2k_\beta^2} v_{\alpha\beta}$$

$$\simeq \sum_\alpha \frac{q^2}{2k_\beta^2} w_{\alpha\beta} \tag{167}$$

which agrees with widely given formulae for the so-called "transport collision frequency" [13]. The collision frequency for weak electron scattering by an array of quasi-stationary statically-screened ions is given, in the Born approximation, by [5]

$$w_{\alpha\beta} = \left(1 - \delta_{\alpha\beta}\right) \frac{2\pi n_i}{\mathfrak{V}} \left| \tilde{V}(\mathbf{q}) \right|^2 S_{ii}(\mathbf{q}) \delta\left( \varepsilon_\alpha - \varepsilon_\beta \right) \tag{168}$$

where $\tilde{V}(\mathbf{q}) = V(\mathbf{q})/\varepsilon_e(\mathbf{q},0)$ is the statically screened potential due to a single ion and $S_{ii}(\mathbf{q})$ is the ion-ion static structure factor. In this approximation, the ions are treated as being infinitely heavy and effectively at rest. Substituting (168) into (167) and transforming the sum over $\alpha$ into an integral, by means of the relation

$$\frac{1}{\mathfrak{V}} = \frac{d^3 \mathbf{k}_\alpha}{(2\pi)^3} = \frac{q k_\alpha}{(2\pi)^2 k_\beta} dq \, dk_\alpha = \frac{m_e}{(2\pi)^2 k_\beta} q \, dq \, d\varepsilon_\alpha \tag{169}$$

which holds in an isotropic medium, yields

$$\nu_\beta = \frac{n_i m_e}{4\pi k_\beta^3} \int_0^{2k_\beta} \left|\tilde{V}(\mathbf{q})\right|^2 S_{ii}(\mathbf{q}) q^3 \, dq$$

(170)

$$= \frac{\langle \nu_0 k^3 \rangle}{k_\beta^3} \ln \Lambda_\beta$$

where $\nu_0(k)$ is the standard (classical) Coulomb collision frequency, as given by (80), and

$$\nu_0 k^3 = \langle \nu_0 k^3 \rangle = 4\pi n_i m_e \left(\frac{Ze^2}{4\pi\varepsilon_0}\right)^2 \tag{171}$$

which is a constant; and

$$\ln \Lambda_\beta = \left(\frac{\varepsilon_0}{Ze^2}\right)^2 \int_0^{2k_\beta} \left|q^2 \tilde{V}(q)\right|^2 S_{ii}(q) \frac{dq}{q} \tag{172}$$

is the Coulomb Logarithm in the Born approximation, in terms of the statically screened potential and the ion static structure factor.

Returning to equation (165) and multiplying both sides by $\mathbf{k}_\beta \tau_\beta$ and summing over $\beta$ yields

$$\sum_\beta k_\beta^2 \tau_\beta p_\beta q_\beta = \tfrac{1}{2} \sum_{\alpha,\beta} \nu_{\alpha\beta} \left(\mathbf{k}_\alpha \tau_\alpha - \mathbf{k}_\beta \tau_\beta\right)^2 \tag{173}$$

and hence

$$\sum_\beta k_\beta^2 \tau_\beta p_\beta q_\beta = \frac{\left(\sum_\beta k_\beta^2 \tau_\beta p_\beta q_\beta\right)^2}{\tfrac{1}{2}\sum_{\alpha,\beta} \nu_{\alpha\beta} \left(\mathbf{k}_\alpha \tau_\alpha - \mathbf{k}_\beta \tau_\beta\right)^2} \tag{174}$$

The significance of this formula is that solving equation (165) in an isotropic medium is equivalent to determining the maximum value of the right hand side of (174) with respect to the parameter set $\{\tau_\beta\}$. This is the basis of a variational approach [24] to solving the Boltzmann equation.

The mean collision frequency, in terms of which the resistivity is expressed by (7), is defined, in the first instance, by

$$\frac{1}{\bar{\nu}} = \frac{2}{3n_e T \mathfrak{V}_e} \sum_\beta \varepsilon_\beta \tau_\beta p_\beta q_\beta \tag{175}$$

which, when combined with (174), yields

$$\overline{V} = \frac{3Tn_e \mathfrak{V}}{2} \frac{m_e \sum_{\alpha,\beta} v_{\alpha\beta} (\mathbf{k}_\alpha \tau_\alpha - \mathbf{k}_\beta \tau_\beta)^2}{\left(\sum_\beta k_\beta^2 \tau_\beta p_\beta q_\beta\right)^2} \quad (176)$$

The right hand side of this equation, is, as has already been noted, is an extremum (minimum) with respect to $\{\tau_\beta\}$. Therefore, if an approximate solution for $\tau_\beta = 1/v_\beta$ is substituted, the overall error will be, at most, second order. Holding to the assumption that the relaxation time remains close to the solution for quasi-elastic scattering, for which, to logarithmic accuracy, $\tau_\beta \propto k_\beta^3$, while making use of equation (173), thus leads to the following approximate result for the linearized form of the mean collision frequency,

$$\overline{V} = \frac{3Tn_e \mathfrak{V}}{2} \frac{2m_e \sum_\beta k_\beta^2 \tau_\beta^2 v_\beta p_\beta q_\beta}{\left(\sum_\beta k_\beta^2 \tau_\beta p_\beta q_\beta\right)^2} \equiv \frac{3Tn_e \mathfrak{V}}{2} \frac{2m_e \sum_\beta k_\beta^2 v_\beta p_\beta q_\beta}{\left(\sum_\beta k_\beta^2 \tau_\beta p_\beta q_\beta\right)\left(\sum_\beta k_\beta^2 \tau_\beta^{-1} p_\beta q_\beta\right)}$$

$$= \frac{3Tn_e \mathfrak{V}}{2} \frac{2m_e \sum_\beta k_\beta^2 v_\beta p_\beta q_\beta}{\left(\sum_\beta k_\beta^5 p_\beta q_\beta\right)\left(\sum_\beta k_\beta^{-1} p_\beta q_\beta\right)} \frac{L_1}{L_2} \quad (177)$$

$$\simeq \frac{3T_B}{2} \frac{\langle\langle \varepsilon v \rangle\rangle}{\langle\langle \varepsilon^{5/2} \rangle\rangle \langle\langle \varepsilon^{-1/2} \rangle\rangle}$$

Where $L_1 = \langle\langle \varepsilon^{5/2} \rangle\rangle / \langle\langle \varepsilon^{5/2} / \ln \Lambda \rangle\rangle$, $L_2 = \langle\langle \varepsilon^{-1/2} \ln \Lambda \rangle\rangle / \langle\langle \varepsilon^{-1/2} \rangle\rangle$ and the approximation amounts to setting $L_1 = L_2$ (which, note, holds exactly in the degenerate limit). The two formulae (175) and (177) are approximately equivalent provided that the relaxation times are (at least approximate) solutions of the Boltzmann equation. The linear form (177) has the advantage of being evaluated near the extremum so that errors are of second order, while it conforms to Mattheissen's rule [24], whereby the resistivity can be linearly decomposed into contributions from different scattering processes.

Another appealing feature of (177) is that the averages $\langle\langle \varepsilon^{5/2} \rangle\rangle$ and $\langle\langle \varepsilon^{-1/2} \rangle\rangle$ appearing in the denominator can be evaluated analytically for a free electron gas. It is straightforward to show that

$$\langle\langle \varepsilon^{-1/2} \rangle\rangle = \frac{T^{-1/2}}{I_{1/2}(\eta)} \ln(1 + e^\eta) \quad (178)$$

while the general relation $\langle\langle \varepsilon^n \rangle\rangle = (n+\tfrac{1}{2})T_B \langle \varepsilon^{n-1} \rangle$ directly yields

$$\langle\langle \varepsilon^{5/2} \rangle\rangle = \frac{3T_B}{(2m_e)^{3/2}} \langle k^3 \rangle \tag{179}$$

## A.2 Application to non-Lorentzian plasmas

The methods described in Appendix A.1 with $\tau_\beta = \tau_\beta^{ei}$, $\nu_{\alpha\beta} = \nu_{\alpha\beta}^{ei}$ etc, apply only to Lorentzian plasmas in which electron-electron (ee) collisions are not explicitly treated. The inclusion of an explicit treatment of electron-electron collisions is not straightforward because of the different ways such collisions affect the transport. Clearly, collisions occurring just among the electrons do not affect their total momentum, and leave the current unchanged. Nevertheless electron-electron collisions do play an important role, in maintaining LTE for example, and neglecting their effect on the transport *in toto* requires justification at the very least.

In the first instance, equation (153) can be broken down into independent ee and ei terms as follows

$$\left.\frac{\partial p_\beta}{\partial t}\right|_{\text{collisions}} = \sum_\alpha \left(\nu_{\beta\alpha}^{ei} - \nu_{\alpha\beta}^{ei}\right) + \sum_\alpha \left(\nu_{\beta\alpha}^{ee} - \nu_{\alpha\beta}^{ee}\right) \tag{180}$$

The corresponding breakdown of equation (154) must take account of the different ways in which the ee and ei collisions act. Collisions among the electrons are presumed to act so as to thermalize the electrons in the moving frame of their center-of-mass (CM) without altering the total momentum. Electron-electron collisions alone therefore make no contribution to the resistivity. Electron-ion collisions, on the other hand, act so as to create an isotropic electron velocity distribution in the ion CM frame (Landau collision integral) or the electron-ion CM frame (Lenard-Balescu collision integral) without affecting the energy distribution. Introducing separate ee and ei relaxation times $\tau_\beta^{ee}$ and $\tau_\beta^{ei}$, the collision term in the Boltzmann equation becomes

$$\left.\frac{\partial p_\beta}{\partial t}\right|_{\text{collisions}} = -\frac{p_\beta - p_\beta^{(0)}}{\tau_\beta^{ei}} - \frac{p_\beta - p^0(\mathbf{k}_\beta - m_e \mathbf{u})}{\tau_\beta^{ee}} \tag{181}$$

where

$$m_e \mathbf{u} = \sum_\beta p_\beta \mathbf{k}_\beta \tag{182}$$

in which **u** is the mean electron drift velocity,

$$\mathbf{u} = \frac{\sigma}{n_e e} \mathbf{E} \tag{183}$$

and

$$p_\beta^{(0)} = \frac{1}{4\pi} \int p_\beta \, \mathrm{d}^2 \hat{\mathbf{k}}_\beta \tag{184}$$

which is the first term in the multipole expansion $p_\beta = \sum_\ell p_\beta^{(\ell)} P_\ell(\cos\theta)$ in terms of $\cos\theta = \hat{\mathbf{k}}_\beta \cdot \hat{\mathbf{u}}$.

Defining the variations $\delta p_\beta$ and $\delta p_\beta^{(\ell)}$ by

$$\delta p_\beta = p_\beta - p_\beta^0 = p_\beta - p^0(\mathbf{k}_\beta)$$

$$\delta p_\beta^{(\ell)} = p_\beta^{(\ell)} : \ell > 0; \quad \delta p_\beta^{(0)} = p_\beta^{(0)} - p_\beta^0 \tag{185}$$

where $\delta p_\beta$ is the total deviation from equilibrium as previously defined, (181) becomes

$$\left. \frac{\partial p_\beta}{\partial t} \right|_{\text{collisions}} = -\frac{p^0(\mathbf{k}_\beta) - p^0(\mathbf{k}_\beta - m_e \mathbf{u})}{\tau_\beta^{ee}} - \frac{\delta p_\beta}{\tau_\beta^0} + \frac{\delta p_\beta^{(0)}}{\tau_\beta^{ei}}$$

$$\simeq \frac{m_e \mathbf{u} \cdot \mathbf{v}_\beta}{\tau_\beta^{ee}} \frac{p_\beta^0 q_\beta^0}{T} - \frac{\delta p_\beta}{\tau_\beta^0} + \frac{\delta p_\beta^{(0)}}{\tau_\beta^{ei}} \tag{186}$$

where

$$\frac{1}{\tau_\beta^0} = \frac{1}{\tau_\beta^{ee}} + \frac{1}{\tau_\beta^{ei}} \tag{187}$$

Substituting (186) into the Boltzmann Equation (151) making use of (152) with $\mathbf{f} = e\mathbf{E}$, and (183), yields the total variation $\delta p_\beta$ as follows

$$\delta p_\beta = \mathbf{f} \cdot \mathbf{v}_\beta \frac{p_\beta^0 q_\beta^0}{T} \left( \tau_\beta^0 + \frac{\tau_\beta^0}{\tau_\beta^{ee}} \frac{m_e \sigma}{n_e e^2} \right) + \frac{\tau_\beta^0}{\tau_\beta^{ei}} \delta p_\beta^{(0)}$$

$$= \tau_\beta \mathbf{f} \cdot \mathbf{v}_\beta \frac{p_\beta^0 q_\beta^0}{T} + \frac{\tau_\beta^0}{\tau_\beta^{ei}} \delta p_\beta^{(0)} \tag{188}$$

where

$$\tau_\beta = \tau_\beta^0 \left( 1 + \frac{\overline{\tau}}{\tau_\beta^{ee}} \right) \tag{189}$$

in which $\overline{\tau} = 1/\overline{\nu}$ is defined by (7) and is given, for Lorentzian plasmas, by (177). Taking the zeroth moment of (188) yields, for finite $\tau_\beta^0$,

$$\delta p_\beta^{(0)} = 0 \tag{190}$$

which is a statement of the LTE requirement that the spectral distribution of the electrons is maintained. Equation (188) thus reduces to (155) but with $\tau_\beta$ given by (189).

The first moment yields the conductivity, as before, in terms of the effective mean collision time,

$$\bar{\tau} = \sum_\beta \lambda_\beta \tau_\beta \tag{191}$$

where

$$\lambda_\beta = \frac{2}{3T}\varepsilon_\beta p_\beta q_\beta, \quad \sum_\beta \lambda_\beta = 1 \tag{192}$$

Substituting for $\tau_\beta$ from (189) and solving for $\bar{\tau}$, yields

$$\bar{\tau} = \frac{\sum_\beta \lambda_\beta \tau_\beta^0}{1 - \sum_\beta \lambda_\beta \frac{\tau_\beta^0}{\tau_\beta^{ee}}} = \frac{\sum_\beta \lambda_\beta \tau_\beta^0}{\sum_\beta \lambda_\beta \frac{\tau_\beta^0}{\tau_\beta^{ei}}} \tag{193}$$

or, in terms of the corresponding collision frequencies,

$$\bar{\nu} = \frac{1 - \sum_\beta \lambda_\beta \frac{\nu_\beta^{ee}}{\nu_\beta^0}}{\sum_\beta \lambda_\beta \frac{1}{\nu_\beta^0}} = \frac{\sum_\beta \lambda_\beta \frac{\nu_\beta^{ei}}{\nu_\beta^0}}{\sum_\beta \lambda_\beta \frac{1}{\nu_\beta^0}} \tag{194}$$

Three notable properties of equation (194) are

(1) If $\nu_\beta^{ei} = 0$, $\forall \beta$ (no electron-ion collisions) then $\bar{\nu} = 0$, which confirms the earlier assertion that electron-electron collisions alone do not give rise to any electrical resistivity;

(2) If $\nu_\beta^{ee} = 0$, $\forall \beta$ (no electron-electron collisions) then one recovers the Lorentzian formula,

$$\bar{\nu} = \bar{\nu}_L \equiv 1 \bigg/ \sum_\beta \frac{\lambda_\beta}{\nu_\beta^{ei}} ;$$

(3) If $\nu_\beta^{ee}/\nu_\beta^{ei} = \text{constant}$ (independent of $\beta$) then $\bar{\nu} = \bar{\nu}_L$. This provides a new criterion for the validity of the Lorentz model, namely that the ee and ei cross-sections have exactly the same dependence on the energy. For Coulomb collisions, this is generally true at high temperatures, when correlations are unimportant. At lower temperatures, differences are due to the Coulomb Logarithm (eg, collective effects) and correlations (exchange and ion-ion correlations).

### A.3 Expansion of the non-Lorentzian correction in powers of $1/Z$

In general, we can write

$$v_\beta^{ei} = Z^2 a_0 L_\beta^{ei} / k_\beta^3$$

$$v_\beta^{ee} = Z a_0 L_\beta^{ee} / k_\beta^3 \tag{195}$$

where $L_\beta^{ei}$ and $L_\beta^{ee}$ incorporate the slowly-varying logarithmic factors, which suggest that the corrections to the Lorentzian conductivity formula may be representable as a power series in $1/Z$. The effective mean collision frequency (194) can be expressed in the following form,

$$\bar{v} = \frac{\sum_\beta \lambda_\beta \dfrac{v_\beta^{ei}}{v_\beta^0}}{\sum_\beta \lambda_\beta \dfrac{v_\beta^{ei}}{v_\beta^0} \dfrac{1}{v_\beta^{ei}}} \tag{196}$$

which corresponds to the Lorentzian model with $\lambda_\beta$ replaced by

$$\tilde{\lambda}_\beta = \lambda_\beta \frac{v_\beta^{ei}}{v_\beta^0} \bigg/ \sum_\beta \lambda_\beta \frac{v_\beta^{ei}}{v_\beta^0} = \lambda_\beta \frac{\tau_\beta^0}{\tau_\beta^{ei}} \bigg/ \sum_\beta \lambda_\beta \frac{\tau_\beta^0}{\tau_\beta^{ei}} \tag{197}$$

Using (195), $\dfrac{v_\beta^{ei}}{v_\beta^0} = \dfrac{1}{1 + b_\beta / Z}$, where $b_\beta = L_\beta^{ee} / L_\beta^{ei}$, which, when substituted into (197), yields

$$\tilde{\lambda}_\beta = \lambda_\beta \left( 1 - \frac{\Delta_\beta^{(1)}}{Z} + \frac{\Delta_\beta^{(2)}}{Z^2} - \frac{\Delta_\beta^{(3)}}{Z^3} + \ldots \right) \tag{198}$$

where

$$\sum_\beta \lambda_\beta \Delta_\beta^{(n)} = 0, \quad n \geq 1 \tag{199}$$

$$\Delta_\beta^{(1)} = b_\beta - \bar{b}, \quad \Delta_\beta^{(n)} = b_\beta^{\,n} - \overline{b^n} - \sum_{r=1}^{n-1} \Delta_\beta^{(r)} \overline{b^{n-r}} \tag{200}$$

$$\overline{b^n} = \sum_\beta \lambda_\beta b_\beta^{\,n} \tag{201}$$

in which the coefficients $\Delta_\beta^{(n)}$ are defined by the series expansion

$$\frac{1+\sum_{n=1}^{\infty}(b_\beta x)^n}{1+\sum_{n=1}^{\infty}\overline{b^n}x^n}=1+\sum_{n=1}^{\infty}\Delta_\beta^{(n)}x^n \tag{202}$$

Then

$$\frac{\overline{\tau}}{\overline{\tau}_L}=\frac{\sum_\beta\tilde{\lambda}_\beta\tau_\beta^{ei}}{\sum_\beta\lambda_\beta\tau_\beta^{ei}}=\frac{\sum_\beta\tilde{\lambda}_\beta k_\beta^3\frac{1}{L_\beta^{ei}}}{\sum_\beta\lambda_\beta k_\beta^3\frac{1}{L_\beta^{ei}}}=\frac{\overline{L^{ei}}}{\widetilde{L^{ei}}}\frac{\sum_\beta\tilde{\lambda}_\beta k_\beta^3}{\sum_\beta\lambda_\beta k_\beta^3} \tag{203}$$

where

$$\overline{L^{ei}}=\frac{\sum_\beta\lambda_\beta k_\beta^3}{\sum_\beta\lambda_\beta k_\beta^3\frac{1}{L_\beta^{ei}}}$$

$$\widetilde{L^{ei}}=\frac{\sum_\beta\tilde{\lambda}_\beta k_\beta^3}{\sum_\beta\tilde{\lambda}_\beta k_\beta^3\frac{1}{L_\beta^{ei}}} \tag{204}$$

The quantities $\overline{L^{ei}}$ and $\widetilde{L^{ei}}$ represent harmonic averages of a slowly varying function with different, but similar, weighting functions, which can therefore be considered to be approximately equal to a degree that any deviation of their ratio from unity can be safely ignored in terms of the effect of their ratio on the individual coefficients in the expansion in powers of $1/Z$. Hence

$$\frac{\overline{\tau}}{\overline{\tau}_L}\simeq\frac{\sum_\beta\tilde{\lambda}_\beta k_\beta^3}{\sum_\beta\lambda_\beta k_\beta^3}=1-\frac{A_1}{Z}+\frac{A_2}{Z^2}-\frac{A_3}{Z^3}+\ldots \tag{205}$$

where, by means of (198), the coefficients are given by

$$A_n=\overline{\overline{\Delta^{(n)}}}\equiv\sum_\beta\lambda_\beta k_\beta^3\Delta_\beta^{(n)}\Big/\sum_\beta\lambda_\beta k_\beta^3 \tag{206}$$

whereupon, using (200),

$$A_1=\overline{\overline{b}}-\overline{b},\qquad A_n=\overline{\overline{b^n}}-\overline{b^n}-\sum_{r=1}^{n-1}A_r\overline{b^{n-r}} \tag{207}$$

where $\overline{b^n}$ is given by (201) and

$$\overline{\overline{b^n}}\equiv\sum_\beta\lambda_\beta k_\beta^3 b_\beta^n\Big/\sum_\beta\lambda_\beta k_\beta^3 \tag{208}$$

Thus the non-vanishing of the coefficients is primarily due to the strong $k^3$ weighting of the averaging (208) of the ratio of the Coulomb logarithms, compared with the standard averaging (201).

# APPENDIX B  MATHEMATICAL THEOREMS AND LEMMAS.

## B.1 The generalized Lenard-Balescu collision integral in the non-degenerate limit

In this appendix we consider the generalization of the linearized Lenard-Balescu collision integral defined by

$$\mathcal{L}_\nu^\mu(T) = \frac{1}{4\pi}\int q^\nu \, d^3\mathbf{q} \int_{-\infty}^{\infty} |V(q,\omega)|^2 \left\langle \varepsilon_\mathbf{k}^\mu \delta(\varepsilon_{\mathbf{k}+\mathbf{q}} - \varepsilon_\mathbf{k} - \omega) \right\rangle_\mathbf{k} \left\langle \delta(E_\mathbf{p} - E_{\mathbf{p}-\mathbf{q}} - \omega) \right\rangle_\mathbf{p} d\omega \quad (209)$$

where $\varepsilon_\mathbf{k} = k^2/2m_e$ and $E_\mathbf{p} = p^2/2m_i$ in the Maxwell-Boltzmann limit of low electron degeneracy, and

$$V(q,\omega) = \frac{Z_i e^2}{q^2 \varepsilon_0 \varepsilon(q,\omega)} \quad (210)$$

is the dynamically screened electron-ion potential. It is further assumed that the electrons and ions are in mutual local thermodynamic equilibrium so that $T_e = T_i = T$. In (209), $\mu$ and $\nu$ are arbitrary real indices of low order.

The argument of the first $\delta$-function is $\mathbf{k}\cdot\mathbf{q}/m_e + q^2/2m_e - \omega$, which yields

$$k_z = \left(\omega - \frac{q^2}{2m_e}\right)\frac{m_e}{q} \quad (211)$$

Hence, for an isotropic Maxwell-Boltzmann distribution of electrons,

$$\left\langle \varepsilon_\mathbf{k}^\mu \delta(\varepsilon_{\mathbf{k}+\mathbf{q}} - \varepsilon_\mathbf{k} - \omega) \right\rangle_\mathbf{k} = \left(\frac{2}{3}\langle \varepsilon_\mathbf{k}\rangle_\mathbf{k} + \frac{k_z^2}{2m_e}\right)^\mu \left\langle \delta(\varepsilon_{\mathbf{k}+\mathbf{q}} - \varepsilon_\mathbf{k} - \omega) \right\rangle_\mathbf{k}$$

$$= \left(T + \frac{1}{4\varepsilon_q}(\omega - \varepsilon_q)^2\right)^\mu \left\langle \delta(\varepsilon_{\mathbf{k}+\mathbf{q}} - \varepsilon_\mathbf{k} - \omega) \right\rangle_\mathbf{k} \quad (212)$$

$$= \left(T + \frac{1}{4\varepsilon_q}(\omega - \varepsilon_q)^2\right)^\mu \sqrt{\frac{1}{4\pi\varepsilon_q T}} \exp\left(-\frac{1}{4\varepsilon_q T}(\omega - \varepsilon_q)^2\right)$$

where use has been made of equation (25) applied to electrons rather than ions.

In a similar fashion,

$$\left\langle \delta(E_\mathbf{p} - E_{\mathbf{p}-\mathbf{q}} - \omega) \right\rangle_\mathbf{p} = \sqrt{\frac{1}{4\pi E_q T}} \exp\left(-\frac{1}{4E_q T}(\omega + E_q)^2\right) \quad (213)$$

and hence

$$\langle\delta(\varepsilon_{\mathbf{k+q}}-\varepsilon_{\mathbf{k}}-\omega)\rangle_{\mathbf{k}}\langle\delta(E_{\mathbf{p}}-E_{\mathbf{p-q}}-\omega)\rangle_{\mathbf{p}} = \frac{\sqrt{m_e m_i}}{2\pi q^2 T}\exp\left(-\frac{\omega^2}{2q^2 v^2}-\frac{q^2}{8mT}\right) \quad (214)$$

where

$$m = \frac{m_e m_i}{m_e + m_i} \quad (215)$$

and

$$v^2 = \frac{T}{m_e + m_i} \quad (216)$$

Substituting in accordance with (212) - (214) into (209) yields

$$\mathcal{L}_\nu^\mu(T) = \frac{\sqrt{m_e m_i}}{8\pi^2 T}\int q^{\nu-2}\,e^{-q^2/8mT}\,d^3\mathbf{q}\int_{-\infty}^{\infty}|V(q,\omega)|^2\,e^{-\omega^2/2q^2 v^2}\,d\omega\left(T+\frac{1}{4\varepsilon_q}(\omega-\varepsilon_q)^2\right)^\mu \quad (217)$$

In particular

$$\mathcal{L}_\nu^0(T) = \frac{\sqrt{m_e m_i}}{8\pi^2 T}\int q^{\nu-2}\,e^{-q^2/8mT}\,d^3\mathbf{q}\int_{-\infty}^{\infty}|V(q,\omega)|^2\,e^{-\omega^2/2q^2 v^2}\,d\omega \quad (218)$$

Now, the exponential factors limit the contributions to the integral to $q^2 \lesssim 8mT$ and $|\omega|\lesssim \sqrt{2}qv \lesssim 4T\sqrt{m/(m_e+m_i)}$ while the integrand is an even function of $\omega$. Expanding the factor $\left(1+\frac{1}{4\varepsilon_q T}(\omega-\varepsilon_q)^2\right)^\mu$ about $\omega=0$ as far as $\mathcal{O}(\omega^2)$ yields, noting that odd powers of $\omega$ do not contribute,

$$\left(1+\frac{1}{4\varepsilon_q T}(\omega-\varepsilon_q)^2\right)^\mu$$

$$= \left(1+\frac{q^2}{8mT}\right)^\mu\left(1+\mu\left(1+\frac{q^2}{8mT}\right)^{-1}\left(\left(\frac{\omega^2}{4\varepsilon_q T}-\frac{\omega}{2T}\right)+\tfrac{1}{2}(\mu-1)\left(1+\frac{q^2}{8mT}\right)^{-1}\left(\frac{\omega}{2T}\right)^2\right)+\ldots\right)$$

$$\lesssim \left(1+\frac{q^2}{8mT}\right)^\mu\left(1+\mathcal{O}(\omega)+\tfrac{1}{2}\mu\left(\frac{\omega}{4T}\right)^2+\mathcal{O}(\omega^3)\right)$$

$$\lesssim \left(1+\frac{q^2}{8mT}\right)^\mu\left(1+\mathcal{O}(\omega)+\tfrac{1}{2}\mu\frac{m}{m_e+m_i}+\mathcal{O}(\omega^3)\right)$$

(219)

Therefore, given that $m_i \gg m_e$, the factor $\left(T + \frac{1}{4\varepsilon_q}(\omega - \varepsilon_q)^2\right)^\mu$ can reasonably be approximated by its value at $\omega = 0$. The result is

$$\mathcal{L}_\nu^\mu(T) \simeq \frac{\sqrt{m_e m_i}}{8\pi^2} T^{\mu-1} \int q^{\nu-2} \left(1 + \frac{q^2}{8m_e T}\right)^\mu e^{-q^2/8mT} d^3\mathbf{q} \int_{-\infty}^\infty |V(q,\omega)|^2 e^{-\omega^2/2q^2 v^2} d\omega \qquad (220)$$

For a Boltzmann electron gas, Boercker et al [19] show that, in an approximation that depends on $m_e \ll m_i$,

$$\int_{-\infty}^\infty \frac{1}{|\varepsilon(q,\omega)|^2} \exp\left(-\frac{1}{2}\left(\frac{\omega}{qv}\right)^2\right) d\omega = \sqrt{2\pi} v q \frac{1}{|\varepsilon_e(q,0)|^2} S_{ii}(q) \qquad (221)$$

(see also section 4.3) and which is trivially exact in the weak-coupling limit when $\varepsilon(q,\omega) = 1$ and $S_{ii}(q) = 1$. This is equivalent to,

$$\int_{-\infty}^\infty |V(q,\omega)|^2 \exp\left(-\frac{1}{2}\left(\frac{\omega}{qv}\right)^2\right) d\omega = \sqrt{2\pi} v q |\tilde{V}(q)|^2 S_{ii}(q) \qquad (222)$$

where $S_{ii}(q)$ is the static ion structure factor, and $\tilde{V}(q) = V(q)/\varepsilon_e(q,0)$. Substitution of these results into (218) - (220) yields the concluding result of this appendix.

$$\mathcal{L}_\nu^\mu(T) = \sqrt{\frac{m}{2\pi}} \frac{T^{\mu-1/2}}{4\pi} \int q^{\nu-1} |\tilde{V}(q)|^2 S_{ii}(q) \left(1 + \frac{q^2}{8m_e T}\right)^\mu e^{-q^2/8mT} d^3\mathbf{q}$$

$$= \sqrt{\frac{m}{2\pi}} T^{\mu-1/2} \int_0^\infty q^{\nu+1} |\tilde{V}(q)|^2 S_{ii}(q) \left(1 + \frac{q^2}{8m_e T}\right)^\mu e^{-q^2/8mT} dq$$

(223)

## APPENDIX C LIST OF SYMBOLS USED FOR MATHEMATICAL AND PHYSICAL QUANTITIES

Note: Throughout this article, Planck's constant, $\hbar$, and Boltzmann's constant, $k_B$, are both set equal to unity. This means that temperature, energy and frequency are all to be considered as given in the same units.

$A_{\mathbf{q}}^e$, $B_{\mathbf{q}}^e$  Coefficients in equation (111).

$A_n$  Coefficients in expansion of $\bar{\tau}/\bar{\tau}_L$ in powers of $1/Z$ as per equation (205).

$a$, $b$  Coefficients, which may be depend on $\mathbf{q}$, in (108).

$a_k$  Electron-ion Rutherford length $= Ze^2/8\pi\varepsilon_0\varepsilon_{\mathbf{k}}$.

$b_\beta$  $= L_\beta^{ee}/L_\beta^{ei} = Zv_\beta^{ee}/v_\beta^{ei}$.

$\overline{b^n}$  $= \sum_\beta \lambda_\beta b_\beta^{\ n}$, $n = 1, 2, 3...$

$c$  Sound velocity.

$B^i(x)$  Function defined by (96).

$\tilde{B}^i(x)$  Function defined by (94).

$D_e$  Degeneracy-modified electron Debye length $= \Omega_e^{-1}\sqrt{T_B/m_e}$.

$D_i$  Ion Debye length $= \Omega_i^{-1}\sqrt{T_i/m_i}$.

$D$  Plasma Debye length $= (1/D_e^2 + 1/D_i^2)^{-1/2}$.

$E_\mathbf{p}$  Energy of ion state with wavevector $\mathbf{p}$, $= \mathbf{p}^2/2m_i$.

$e$  Unit of electronic charge.

$F(x)$  Function defined by (121).

$F_0(\mathbf{q})$  Coefficient in the modal decomposition of the ion structure factor (47).

$G$  $= (\Omega_{\mathbf{q}}^e/\Omega_e)^2$.

$g$  Spin degeneracy factor.

$g_{\mathbf{q}}^e(\omega)$  Collective part of the ion structure factor, (47).

$I_j(\eta)$    Fermi integral defined by $I_j(\eta) = \int_0^\infty \dfrac{y^j}{1+\exp(y-\eta)} dy$

$I_\mathbf{q}$    Electron-ion momentum exchange integral defined by (102).

$I_\mathbf{q}^{\text{plasmon}}$    Electron plasmon contribution to $I_\mathbf{q}$.

$\mathcal{J}_n(z)$    Debye integral, (79).

$K$    Upper wavenumber limit for ion-acoustic modes.

$K_j(x)$    Modified Bessel function of the second kind [27].

$\mathbf{k}$    Electron wavevector.

$k_F$    Fermi momentum $= (3\pi^2 n_e)^{1/3}$.

$L$    Landau length $= Z^2 e^2/4\pi\varepsilon_0 T$.

$L(u,v;\eta)$    Function defined by (24), which provides an exact algebraic representation of the dynamic structure factor for non-interacting fermions.

$L_\beta^{\text{ei}}$    Coulomb logarithm for electron-ion collisions given by (195).

$L_\beta^{\text{ee}}$    Coulomb logarithm for electron-electron collisions given by (195).

$\overline{L^{\text{ei}}}$, $\widetilde{L^{\text{ei}}}$    Harmonic averages of $L_\beta^{\text{ei}}$ as defined by (204).

$\mathfrak{L}_\nu^\mu(T)$    Integral defined by (209).

$m$    Mass. Electron – ion reduced mass.

$m_e$    Electron mass.

$m_i$    Ion mass.

$\mathcal{N}_\mathbf{q}$    Bose-Einstein function $= \left(\exp(\Omega_\mathbf{q}/T)-1\right)^{-1}$.

$n_e$    Electron density.

$n_i$    Ion density.

$\mathbf{p}$    Ion momentum/wavevector.

$p_\beta$    Average occupancy of electron state $\beta$, $= p(\mathbf{k}_\beta)$

$p_\beta$    (Main text) = Fermi-Dirac distribution, $p_\beta = p(\varepsilon_\beta) = \left(1+\exp(\varepsilon_\beta/T_e - \eta)\right)^{-1}$

$p_\beta^0$    LTE electron distribution (where formally distinct from $p_\beta$) = Fermi-Dirac distribution,

$$p_\beta^0 = p^0(\varepsilon_\beta) = \left(1 + \exp(\varepsilon_\beta/T_e - \eta)\right)^{-1}$$

$p_\beta^{(\ell)}$    Coefficient of $\ell^{\text{th}}$ order term in multipole expansion of $p(\mathbf{k}_\beta)$

**q**    General wavevector variable. Momentum transfer in electron-ion collision.

$q$    $1 - p$.

$R_0$    $= I'_{1/2}(\eta)/I_{1/2}(\eta) = T/T_B$

$S(\mathbf{q}, \omega)$ Dynamic structure factor.

$S(\mathbf{q})$    Static structure factor.

$S_{ee}^0(\mathbf{q}, \omega)$ Non-interacting dynamic structure factor for electrons, given by (24).

$S_{ii}^0(\mathbf{q}, \omega)$ Non-interacting dynamic structure factor for ions, given by (25).

$S_{ee}(\mathbf{q}, \omega)$ Interacting dynamic structure factor for electrons.

$S_{ii}(\mathbf{q}, \omega)$ Interacting dynamic structure factor for ions.

$\tilde{S}_{ii}(\mathbf{q}, \omega)$ Interacting dynamic structure factor for ions, taking account the modifying effect of the electrons.

$S_{ii}(\mathbf{q})$ Static structure factor for ions.

$\tilde{S}_{ii}(\mathbf{q})$ Static structure factor for ions, taking account the modifying effect of the electrons.

$S_0(\mathbf{q})$ Classical ion static structure factor ($T \gg \Omega_i$).

$T$    Temperature

$T_B$    Effective temperature or energy that is equal to the electronic bulk modulus divided by the electron density $= T_e/\langle q \rangle = T_e\, I_{1/2}(\eta)/I'_{1/2}(\eta)$

$T_e$    Electron temperature.

$T_F$    Fermi energy/temperature.

$T_i$    Ion temperature.

**u**    Electron drift velocity.

$u$    $= \omega/T_e$

$\mathfrak{V}$    Volume.

$V(\mathbf{q})$    Electron-ion potential, taken to be Coulomb ($= Ze^2/\varepsilon_0 q^2$).

$\tilde{V}(\mathbf{q})$    Statically-screened electron-ion potential $= V(\mathbf{q})/\varepsilon_e(\mathbf{q},0)$.

$V(\mathbf{q},\omega)$ Dynamically-screened electron-ion potential $= V(\mathbf{q})/\varepsilon(\mathbf{q},\omega)$.

$\boldsymbol{v}$      Electron velocity $= \mathbf{k}/m_e$.

$v$      Mean thermal velocity of ion or electron-ion system, $= T/m_i$ or $T/(m_e + m_i)$

$v_F$      Fermi velocity.

$v$      $= q^2/2m_e T_e$

$x$      $= q^2 D_e^2$

$x_0$      Value of $x$ corresponding to the saddle point of $\Delta\Omega_i^e$ in the treatment of the integral (124).

$y$      $= q^2/2m_e \Omega^i(x)$

$Z$      Effective ion charge.

$\alpha, \beta$    Electron state labels.

$\gamma_i$      Ion acoustic mode "adiabaticity parameter" as defined in equation (57).

$\gamma(x) = \sqrt{G(x)}$

$\gamma_0$    $= \gamma(x_0) = \sqrt{G(x_0)}$

$\gamma_e(Z)$ Spitzer's correction factor for correcting for ee collisions, as defined by (141).

$\Delta$      $= \Delta\Omega_e^i = \dfrac{(1-y)^2}{4y}\Omega^i(x) = \dfrac{m_e}{2q^2}\left(\Omega_\mathbf{q}^i - \dfrac{q^2}{2m_e}\right)^2$

$\Delta_0$    $= \varepsilon_{q/2} = q^2/8m_e = x/8m_e D_e^2 = \lim\limits_{y\to\infty}\Delta$

$\Delta_\beta^{(n)}$   Coefficients in expansion of $\tilde{\lambda}_\beta$ about $\lambda_\beta$ in powers of $1/Z$, as per equation (198).

$\Delta\Omega_e^i$   $= \dfrac{(1-y)^2}{4y}\Omega^i(x) = \dfrac{m_e}{2q^2}\left(\Omega_\mathbf{q}^i - \dfrac{q^2}{2m_e}\right)^2.$

$\Delta\Omega_i^e$   Function defined by (118), $= \dfrac{m_i}{2q^2}\left(\Omega_\mathbf{q}^e - \dfrac{q^2}{2m_i}\right)^2.$

$\delta(x)$    Delta function.

$\delta_\mathbf{q}(\omega)$    Rayleigh line profile part of the ion dynamic structure factor as defined by (47) - (48).

$\delta p_\beta$    $= p_\beta - p_\beta^0 =$ total deviation of the electron distribution $p_\beta$ from equilibrium.

$\delta p_\beta^{(\ell)}$    Coefficient of $\ell^{th}$ order term in multipole expansion of $\delta p_\beta$.

$\varepsilon$    Electronic energy.

$\varepsilon_\mathbf{k}$    Energy of electron quasiparticle state with wavevector $\mathbf{k}$.

$\varepsilon_F$    Fermi energy $= k_F^2/2m_e = \left(\tfrac{3}{2} I_{1/2}(\eta)\right)^{2/3} T_e$.

$\varepsilon_0$    Permittivity of free space.

$\varepsilon(\mathbf{q},\omega)$ Longitudinal dielectric function.

$\varepsilon_e(\mathbf{q},\omega)$ Longitudinal dielectric function for electrons.

$\varepsilon_e^0(\mathbf{q},\omega)$    Longitudinal dielectric function of a non-interacting Fermi gas.

$\tilde{\varepsilon}_i(\mathbf{q},\omega)$ Longitudinal dielectric function for ions, taking account of the modifying effect of the electrons.

$\eta$    Electron degeneracy parameter, $= \mu/T_e$, determined by $I_{1/2}(\eta) = \tfrac{2}{3}(\varepsilon_F/T_e)^{3/2}$.

$\overline{\Lambda}$    Argument of Coulomb logarithm $\ln(\overline{\Lambda})$ defined by (82).

$\overline{\Lambda}_0$    Argument of classical Coulomb logarithm, as given by (85).

$\Lambda^{plasmon}$ Electron collective contribution to the thermally averaged Coulomb logarithm.

$\lambda$    $= D_i^2/D_e^2$.

$\lambda_\beta$    Normalised weighting coefficient defined by (192).

$\tilde{\lambda}_\beta$    Normalised weighting coefficient defined by (197)

$\mu$    Electronic chemical potential.

$\nu(k)$    Effective collision frequency for electrons with momentum $k$, $=1/\tau(k)$.

$\nu_F$    Effective collision frequency for electrons at the Fermi energy, $=\nu(k_F)=1/\tau_F$.

$\nu_\beta$    $= \nu(k_\beta)$

$\nu_\beta^{ei}$    Effective electron-ion collision frequency for electrons with momentum $k_\beta$, $=1/\tau_\beta^{ei}$

| | |
|---|---|
| $\nu_\beta^{ee}$ | Effective electron-electron collision frequency for electrons with momentum $k_\beta$, $=1/\tau_\beta^{ee}$ |
| $\nu_\beta^0$ | $= \nu_\beta^{ei} + \nu_\beta^{ee}$ |
| $\bar{\nu}$ | Mean collision frequency, defined by (7). |
| $\bar{\nu}_L$ | Mean collision frequency as given by the Lorentzian model, $=1 / \sum_\beta \lambda_\beta / \nu_\beta^{ei}$ |
| $\nu_0(k)$ | Classical collision frequency, (80), used in the definition of the Coulomb logarithm. |
| $\nu_c$ | $= \bar{\nu}_0 =$ standard collision frequency defined by (81). |
| $\nu_{\alpha\beta}$ | Rate of electronic transitions $\beta \to \alpha$ induced by electron-ion collisions (Lorentzian model). |
| $\nu_{\alpha\beta}^{ee}$ | Rate of electronic transitions $\beta \to \alpha$ induced by electron-electron collisions. |
| $\nu_{\alpha\beta}^{ei}$ | Rate of electronic transitions $\beta \to \alpha$ induced by electron-ion collisions (non-Lorentzian model) |
| $\Theta$ | Ion acoustic Debye temperature. |
| $\Theta_D$ | Debye temperature (pertaining to the specific heat of a normal solid). |
| $\Theta_R$ | Effective Debye temperature in the Bloch formula for the resistivity of a metallic solid. |
| $\sigma$ | Electrical conductivity. |
| $\tau(k)$ | Relaxation time, $=1/\nu(k)$. |
| $\bar{\tau}$ | Mean collision time, $=1/\bar{\nu}$ |
| $\bar{\tau}_L$ | Mean collision time in the Lorentzian model $=1/\bar{\nu}_L = \sum_\beta \lambda_\beta \tau_\beta^{ei}$ |
| $\tau_\beta^{ei}$ | Relaxation time for electron-ion collisions $=1/\nu_\beta^{ei}$. |
| $\tau_\beta^{ee}$ | Relaxation time for electron-electron collisions, $=1/\nu_\beta^{ee}$. |
| $\tau_F$ | Relaxation time at the Fermi energy, $=\tau(k_F)=1/\nu_F$. |
| $\tau_\beta$ | Effective relaxation rate for electrons in state $\beta$ = coefficient in the dipole term in the variation of the distribution function as per equations (155) and (188). |
| $\tau_\beta^0$ | $= \left(1/\tau_\beta^{ee} + 1/\tau_\beta^{ei}\right)^{-1} = 1/\nu_\beta^0$ |
| $\Omega_e$ | Electron plasma frequency. |
| $\Omega_i$ | Ion plasma frequency. |
| $\Omega_q^e$ | Frequency of electron (Langmuir) mode with wavevector $\mathbf{q}$. |

$\Omega_{\mathbf{q}}^{i}$     Frequency if ion (or ion acoustic) mode with wavevector **q**.

$\Omega^{i}(x) = \Omega_{\mathbf{q}}^{i}$ expressed in terms of $x = q^2 D_e^2$

$\omega$     Frequency variable representing the energy transfer in a collision.

$\hat{\omega}_e$     Frequency range of the electron dynamic structure factor.

$\hat{\omega}_i$     Frequency range of the ion dynamic structure factor.

$\langle X \rangle$    Denotes the thermal average, $\sum_{\beta} p_{\beta} X_{\beta} \Big/ \sum_{\beta} p_{\beta} = \dfrac{1}{n\mathfrak{V}} \sum_{\beta} p_{\beta} X_{\beta}$.

$\langle X \rangle_{\mathbf{k}}$    Denotes the thermal average, $\sum_{\mathbf{k}} p(\mathbf{k}) X(\mathbf{k}) \Big/ \sum_{\mathbf{k}} p(\mathbf{k})$

$\langle\langle X \rangle\rangle$    Denotes the average, $\sum_{\beta} p_{\beta} q_{\beta} X_{\beta} \Big/ \sum_{\beta} p_{\beta} q_{\beta} = \langle q X \rangle / \langle q \rangle = \dfrac{T_B}{T} \langle q X \rangle$.

$\overline{X} \quad = \sum_{\beta} \lambda_{\beta} X_{\beta}$

$\overline{\overline{X}} \quad = \sum_{\beta} \lambda_{\beta} k_{\beta}^{3} X_{\beta} \Big/ \sum_{\beta} \lambda_{\beta} k_{\beta}^{3}$

This page is intentionally left blank